\documentclass[%
reprint, twocolumn, superscriptaddress,
showpacs,preprintnumbers,
nofootinbib,
 amsmath,amssymb,
 aps,
pra,
]{revtex4}
\usepackage{graphicx}
\expandafter\let\csname equation*\endcsname\relax
\expandafter\let\csname endequation*\endcsname\relax
\usepackage{amsmath}
\usepackage{amssymb}
\usepackage{color}
\newcommand {\fabs}[1] {\left| #1 \right|}

\newcommand{\ket}[1]{\ensuremath{|#1\rangle}}
\newcommand{\bra}[1]{\langle#1|}
\newcommand{\braket}[2]{\langle#1|#2\rangle}

\DeclareMathOperator{\sinc}{sinc}

\usepackage{ulem}

\begin{document}
\title{Selective population of a large-angular-momentum state in an optical lattice}
\author{A. Kiely}
\email{anthony.kiely@ucc.ie}
\affiliation{Departamento de Qu\'{\i}mica F\'{\i}sica, UPV/EHU, Apdo
644, 48080 Bilbao, Spain}
\affiliation{Department of Physics, University College Cork, Cork, Ireland}

\author{J. G. Muga}
\affiliation{Departamento de Qu\'{\i}mica F\'{\i}sica, UPV/EHU, Apdo
644, 48080 Bilbao, Spain}

\author{A. Ruschhaupt}
\affiliation{Department of Physics, University College Cork, Cork, Ireland}
%
%
\begin{abstract}
We propose a method to selectively populate a large angular momentum state of ultracold atoms (each with an orbital angular momentum $l \approx 2 \hbar$) in the Mott regime of a two-dimensional optical lattice. This is done by periodically modulating the lattice amplitude and implementing an additional rotated rectangular lattice of shorter wavelength. The specific pulse sequences are designed using a four--level model for each well and are implemented sequentially. The results are confirmed with numerical simulations of the full Schr\"{o}dinger equation. These methods are another step in constructing a modular toolbox of operations for creating higher orbital states in optical lattices.
\end{abstract}
\maketitle

\section{Introduction}
Optical lattices are periodic potentials formed by interfering monochromatic laser beams, which can trap many ultracold atoms in large arrays \cite{bloch_2008,lewenstein_2012}.  They have found applications in building atomic clocks \cite{takamoto2005} and as a possible architecture for quantum computing \cite{kay2004,calarco2004,shimizu2004,lewen2007}. It has even been made possible to perform single site addressing with the invention of the quantum-gas microscope \cite{bakr2009,sherson2010}. Detailed reviews of quantum gases in
optical lattices can be found in \cite{lewen2007,jaksch1998,bloch2005}. 

They are predicted to be useful quantum simulators for condensed matter physics since they are highly controllable, i.e., one can easily adjust both the periodicity, depth and dimensionality of the potential. A particular milestone in investigating quantum many body physics was the the observation of the phase
transition between a superfluid and a Mott--insulator state
\cite{mott_fluid,mott_fluid_cs}. 

For bosonic atoms, the ground states possible in optical lattices are necessarily positive definite, which is a general property of bosonic ground state wave functions \cite{wu_2009}. However using the orbital degrees of freedom in higher Bloch bands (which have complex nodal geometries), one can explain many complex phases \cite{Ma_2017} and mimic the orbital physics of  electronic matter, e.g., transition metal oxides \cite{tokura_2000,maekawa_2004}. Hence, there has been much interest in studying the effects of higher bands of optical lattices \cite{lewenstein_2011, higher_bands_review}, e.g., extending bosonic Hubbard model to include higher Bloch bands \cite{isacsson_2005} and examining exotic phases arising from the interplay of interactions and the higher bands \cite{hebert_2013}. Experiments have been performed realizing multiorbital systems with ultracold atoms \cite{browaeys_2005, mueller_2007,kock2016,Zhai_2013,Hu_2017} where the lifetimes of atoms were several tunneling times. Properties of atoms loaded in the higher states have been examined theoretically in \cite{sowinski_2013,pinheiro_2015}. 

Engineering quantum states in higher bands is therefore clearly of large interest and several techniques have been developed to manipulate the orbital state of atoms in optical
lattices \cite{higher_bands_review}. The idea of oscillating the lattice position or the lattice depth was first investigated and utilized in \cite{Denschlag_2002}.

Shaking a lattice in one direction (i.e. a periodic modulation of the position of the trap minima) was initially used for renormalizing the tunneling rate \cite{Gemelke2005,Kierig2008}. This allows for dynamical control over the Mott--insulator/ superfluid transition \cite{tunnelling_renorm,tunnelling_renorm_exp} and has also been used to realize the Haldane \cite{Haldane1988,Jotzu2014} and Hofstadter \cite{Hofstadter1976,Aidelsburger2011} models. However it has also been proposed to prepare higher orbital states \cite{zhang_2014,  strater_2015, zhang_2015,kiely_2016,keles_2017} with applications to quantum computation \cite{schneider_2012}, and interferometry of condensates \cite{frank_2014} and non-interacting ultracold atoms \cite{weidner_2017,weidner_2018}. Shaking a lattice has also been implemented experimentally \cite{parker_2013, khamehchi_2016}.

Periodic modulation of the lattice amplitudes has been used in order to induce controlled transitions to higher orbital states \cite{sowinski_2012,lacki_2013}, e.g., creating a cluster of bipartite entangled atom pairs in an optical superlattice \cite{cao_2017} and for the purpose of spectroscopy of the excitation spectrum \cite{stoferle_2004,Kollath_2006,pedersen_2013}. Polychromatic amplitude modulation has also been shown to enhance transport in an optical lattice \cite{Pepino_2016}.

In \cite{kiely_2016}, a four--level model of the motional states of an atom was used to design a protocol of shaking the lattice and varying the interference term in order to create a staggered state of atoms each with angular momentum $l \approx \pm\hbar$ \cite{collin_2010}. A four-band effective Hamiltonian was also used to describe interacting fermions in a shaken square lattice \cite{keles2017}.

In this paper, we wish to extend these methods to create a similar state which has the same angular momentum per atom. Our target state is a complex state which consists of each potential well occupied by a single atom, carrying angular momentum of $\approx 2 \hbar$ (see Fig. \ref{fig_setup_diagram}). By comparison, this state has a large total orbital angular momentum ($\approx 2 N \hbar$ for $N$ particles) since the magnitude of the total angular momentum in the previous case is maximally $\approx \hbar$.

In particular, we propose a method which, starting from a Mott--insulator ground state, prepares such a target state only by dynamically modulating lattice amplitudes. By restricting to the case of single site occupation it has the advantage that heating due to collisions between several oscillating atoms in a single site is avoided in our scheme. Specifically, the atoms are first excited by amplitude modulation. Then in a second step, angular  momentum is transferred to them using an additional rotated lattice. The methods proposed here could also be used together with the results from \cite{kiely_2016} to form a modular system (or building blocks) for creating different higher orbital states.

In \cite{sowinski_2012}, periodic modulation of the lattice amplitudes is used in order to induce controlled transitions to higher orbital states. However in that work, a filling factor of two is assumed and the contact interaction strength between the particles plays an important role. This work differs from the results in \cite{sowinski_2012} as we assume a filling factor of one and use no interaction effects in order to generate the state. The methods presented here are intended to complement those in \cite{kiely_2016}, constructing a modular toolbox of operations for creating higher orbital states in optical lattices.  The use of atoms with angular momentum in an optical lattice has been explored in \cite{pelgri_2017,pietra_2012}. Creating systems of interacting rotating ultra-cold atoms in optical lattices could prove useful for investigating quantum Hall effects \cite{gemelke_2010}. Instead of applying the results to an optical lattice, the required potentials could also be produced by using optical tweezers \cite{beugnon_2007,couvert_2008,kaufman_2013,Carpentier_2013,nogrette_2014,barredo_2016,stuart_2018}.

The remainder of this paper is structured as follows. In the next
section, we derive our approximate model for the optical lattice. In Section \ref{schemes}, we used this model to construct a sequential scheme which prepares the target angular momentum state, using effective fixed area pulses coupling the states. In Section \ref{Numerics}, we perform
numerical simulation of the full Schr\"{o}dinger equation for a single atom in one site of
an optical lattice in order to substantiate the assumptions of our model. In Section \ref{exp_sec}, we comment on the experimental parameter values required. Finally in Section \ref{discuss}, we summarize our results and discuss future extensions.

\section{Model \label{model}}

In this section, we will first present the physical model in detail. Then, a four-level approximation of this setting is derived which will later
allow us to design the required scheme to achieve the target state.

\subsection{Optical lattice\label{optical}}
We consider a two--dimensional optical lattice (in the $x$--$y$ plane) generated by two pairs of counter--propagating laser beams (which we will call the primary lattice). We assume a strong confinement in the $z$ direction such that only dynamics in the $x$--$y$ plane are relevant. This is implemented experimentally by a simple harmonic confinement in the $z$-direction, with a trapping frequency much greater than the other directions (see \cite{bloch_2008,mott_fluid,mott_fluid_cs} for example).
This primary lattice should have a wavelength $\lambda=2 \pi/k$.

\begin{figure}[t]
\begin{center}
\includegraphics[angle=0,width=0.9\linewidth]{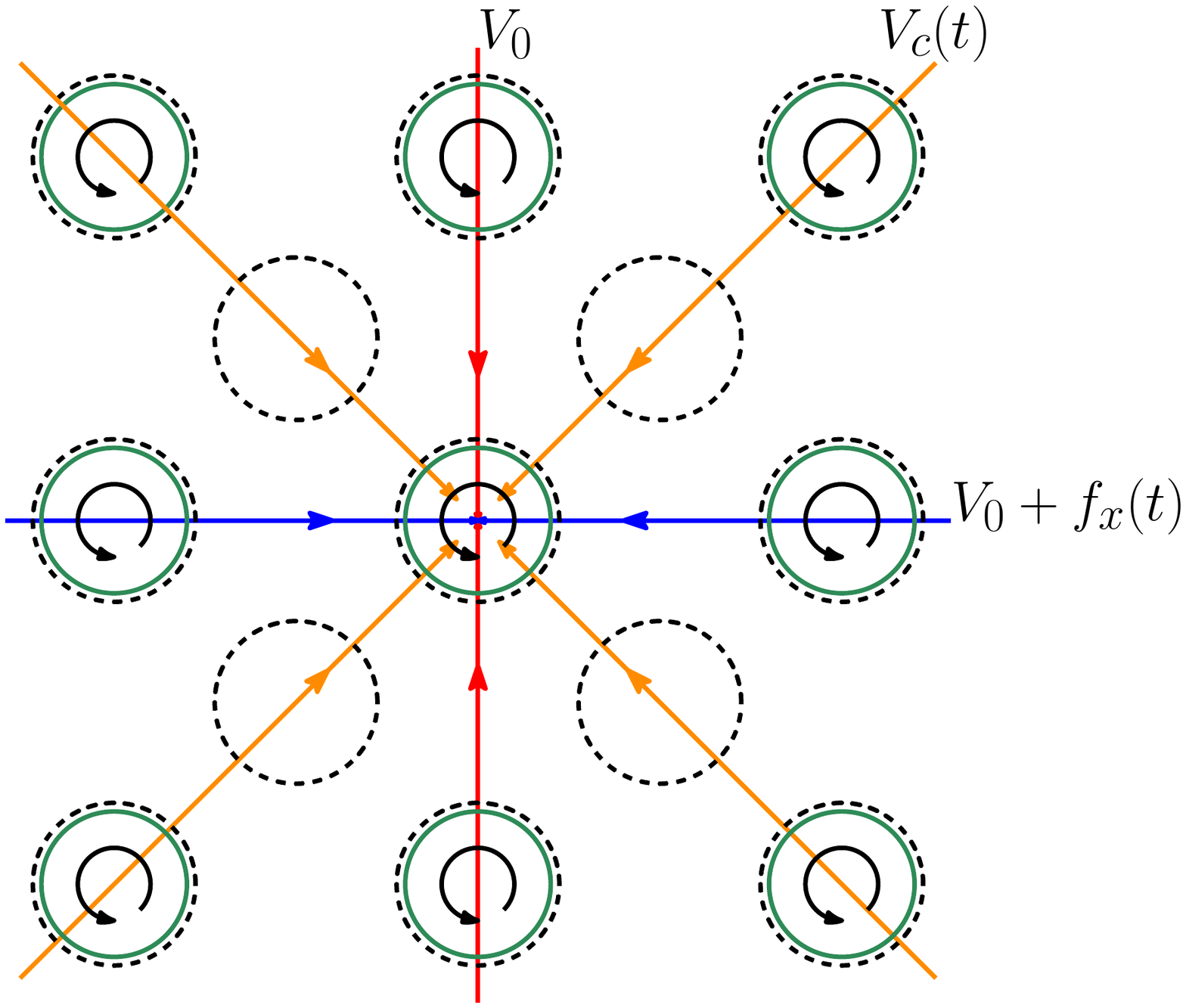}
\end{center}
\caption{Diagram of the counter propagating incident laser beams creating the two modulated lattices. The primary lattice is created by beams in the $x$-direction (blue horizontal lines) and $y$-direction (red vertical lines); the corresponding lattice sites are indicated by green, solid circles.
The rotated, temporary lattice is created by the additional beams (orange diagonal lines) of shorter wavelength at an angle of $\pi/4$ relative to the primary lattice; the corresponding lattice sites are indicated by black, dashed circles. In the target state, each site of the primary lattice contains one atom in state $\ket{+}$ with angular momentum  $\approx 2 \hbar$ (indicated by solid black arcs).\label{fig_setup_diagram}}  
\end{figure}

The Hamiltonian for this lattice alone is separable in $x$ and $y$ and therefore unable to couple the $x$ and $y$ degrees of freedom, which is necessary to generate angular momentum. On account of this, there is an additional rectangular lattice at an angle of $\pi/4$ relative to the primary lattice whose intensity can be varied in time (see diagram in Fig. \ref{fig_setup_diagram}). This will be referred to as the rotated lattice and it is used to transfer angular momentum to the atoms during the preparation scheme of the target state. The rotated lattice is only temporary as it is switched off initially and also again when the preparation of the target state is completed. The rotated lattice has a shorter wavelength $\lambda_{s}=2 \pi/k_{s} = \lambda/\sqrt{2}$ with $k_s = \sqrt{2} k$. Hence, there is always a well of the rotated lattice at each well of the primary lattice, as shown in Fig. \ref{fig_setup_diagram}.

Before continuing, we note that there may be alternative ways to implement the resulting potential, other than optical lattices. One such possibility is optical tweezers which have been previously used for transporting atoms \cite{beugnon_2007, couvert_2008}. Cooling of a single atom to its quantum ground state  \cite{kaufman_2013} and preparation of a single atom in an optical microtrap with high fidelity has been shown \cite{Carpentier_2013}. Even two-dimensional arrays of microtraps with arbitrary geometries \cite{nogrette_2014} and reconfigurable arrays of optical tweezers have been demonstrated for single atoms \cite{stuart_2018}. Optical microtraps could be alternatively used to implement the resulting potential below for a single atom.

Returning to the optical lattice setting, we also assume that the atoms are in the Mott insulator regime with filling factor of one, i.e., each site is occupied by a single atom which is essentially independent of all the others. One can ensure such a regime by having a large lattice amplitude so that tunneling rates are negligible. While it is sufficient to consider each atom separately in the following, it is important to note that the all operations presented here are global and will affect all the atoms/sites simultaneously.

The potential of the primary and the rotated lattices together is given by
\begin{eqnarray}
\lefteqn{V(x,y) = \left[V_{0}+f_{x}(t)\right] \sin^{2}\left(k x\right)+V_{0}\sin^{2}\left(k y\right)} && \nonumber \\
          && + V_{c}(t)\left [\sin^{2}\left(k_{s}\frac{x+y}{\sqrt{2}}\right)+\sin^{2}\left(k_{s}\frac{x-y}{\sqrt{2}}\right)\right ] \\
&=& \left[V_{0}+f_{x}(t)\right] \sin^{2}\left(k x\right)+V_{0}\sin^{2}\left(k y\right) \nonumber \\
          &&-V_{c}(t)\cos\left(2 k x\right)\cos\left(2 k y\right) + V_c (t),
\label{Vpot}
\end{eqnarray}
where $V_{0}+f_{x}(t)$ is the time dependent lattice amplitude in $x$ direction of the primary lattice and $V_{c}(t)$ is the time-dependent amplitude of the rotated lattice potential. We will ignore the time dependent energy shift of $V_{c}(t)$ in \eqref{Vpot} in the following. Note that we assume that the lasers of the unperturbed lattice are blue shifted (i.e. $V_{0}>0$) and we design the protocol so that $V_c \geq 0$ during the process to avoid any problems with the wells becoming too shallow. We also assume that there is no significant interference terms so the potentials simply add up. This could be achieved, for example, by orthogonal polarizations of the lasers or different detunings that cause a rapid time-dependent interference that averages out on the scale of the atomic motion \cite{han_2001}.

The single particle Hamiltonian is given by
\begin{eqnarray}
H(t) &=& H_{0}+H_{1}(t),\label{H_lattice_frame}\\
H_{0} &=& -\frac{\hbar^{2}}{2m}\nabla^{2}+V_0\sin^{2}(k x)+V_0\sin^{2}(k y),\\
H_{1}(t) &=& f_{x}(t)\sin^{2}(k x)-V_{c}(t)\cos\left(2 k x\right)\cos\left(2 k y\right). \label{H_1} \nonumber \\
\end{eqnarray}
The main goal is to design  control schemes, i.e., the time dependence of the functions $f_{x}(t)$ and $V_{c}(t)$, which lead to the desired final state.

More specifically, the amplitude modulation presented here can create two excitations in a given spatial direction. The position modulation (or shaking) outlined in \cite{kiely_2016} can create one excitation in a given spatial direction. In both cases, the part of the wavefunctions in the orthogonal direction must have the same parity for the coupling to be non-zero. In order to couple degenerate states, one can use the $\cos(x)\cos(y)$ type term arising from an extra lattice (see Eq. \eqref{Vpot}), to couple degenerate states which have even-even or odd-odd parity. If the states have an even-odd parity, one can then use the $\sin(x)\sin(y)$ type term arising from a difference in polarization of the laser beams (see \cite{kiely_2016}).

\begin{figure}[t]
\begin{center}
\includegraphics[angle=0,width=0.8\linewidth]{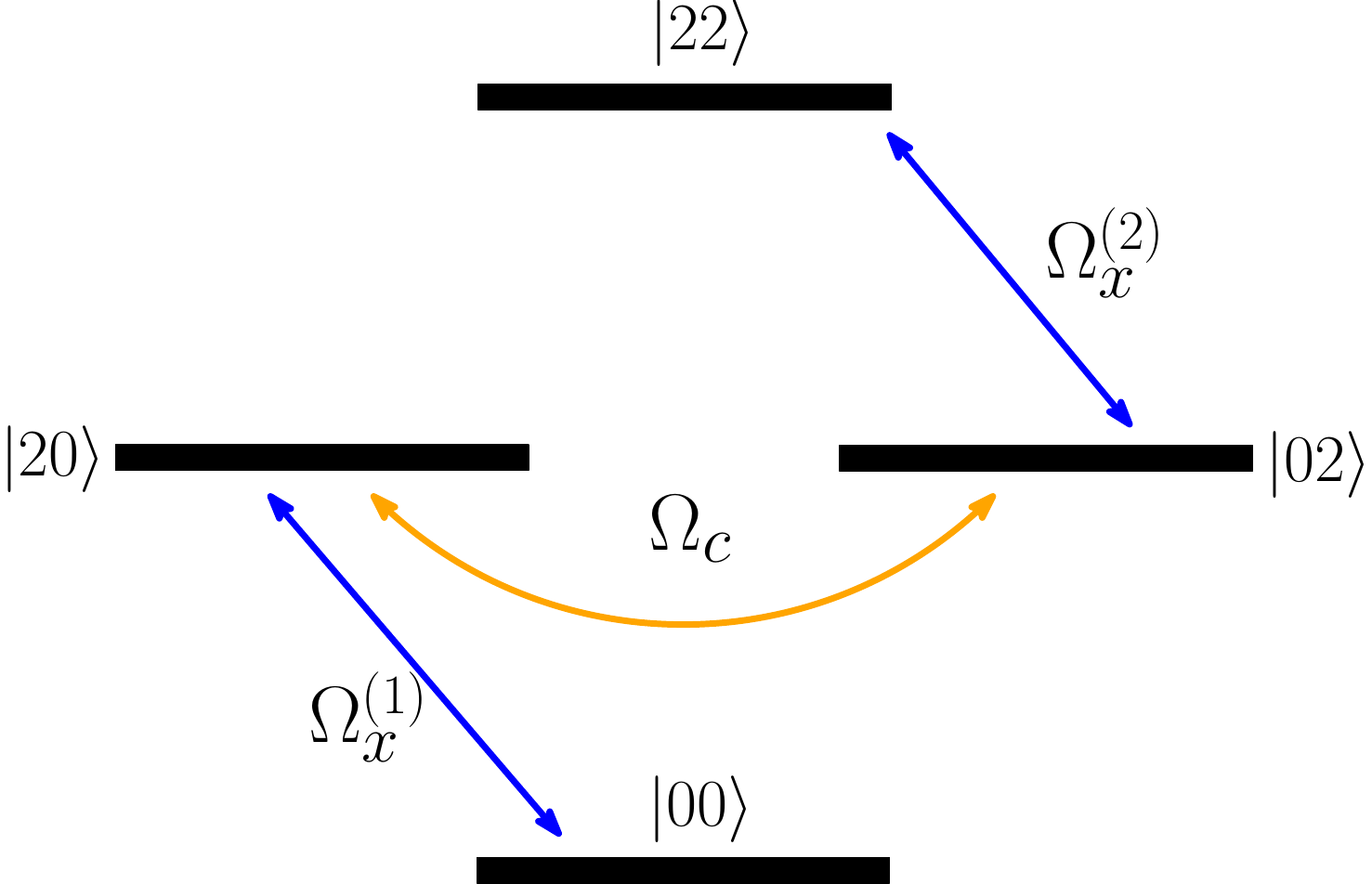}

\end{center}
\caption{Energy level diagram for the four chosen energy eigenstates of $H_{0}$ and the various couplings between them.
 \label{fig_level}}  
\end{figure}

\subsection{Four--level approximation \label{four_level_approx}}

We focus on an individual atom in a single well of the lattice region defined by
$- \ell\le x \le \ell$ and $-\ell\le y \le \ell$, where $\ell=\lambda/4$ is the lattice constant. Interaction effects of other atoms are neglected as we are in the Mott insulator regime with unit filling. 

\begin{figure}[t]
\begin{center}
\includegraphics[angle=0,width=0.95\linewidth]{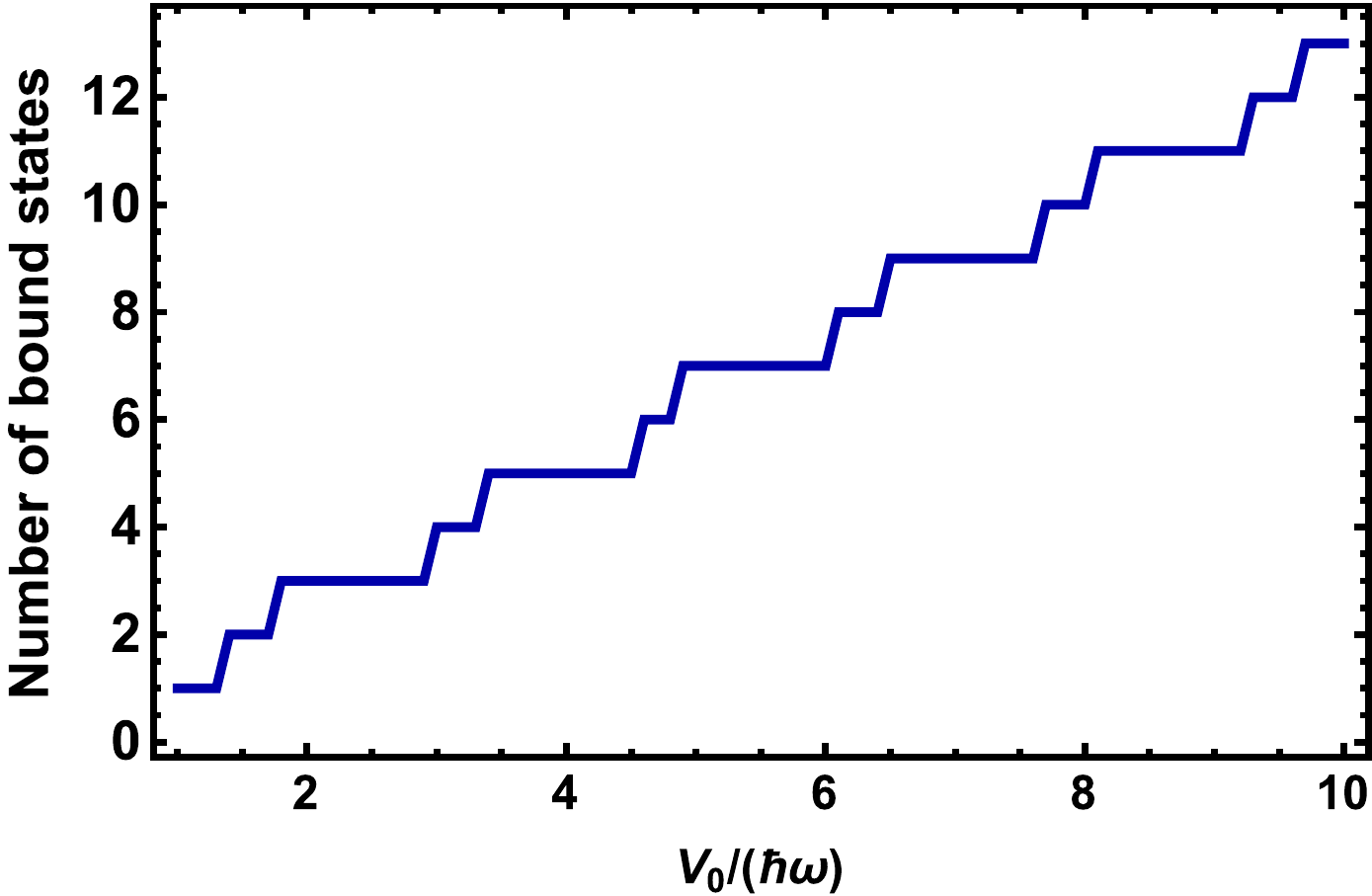}

\end{center}
\caption{Number of bound states in one dimension against lattice depth  $V_{0}$.
 \label{fig_bound}}  
\end{figure}

Analogous to \cite{kiely_2016}, we make a four--level approximation assuming that it is sufficient to considerer only the four most relevant eigenstates of $H_{0}$ localized in the central site. Different from \cite{kiely_2016}, these four eigenstates are now
$\left\{\ket{00},\ket{20},\ket{02},\ket{22}\right\}$ (see Fig. \ref{fig_level});
in coordinate representation, these four basis states are given by $\braket{\vec{r}}{ij}=\Gamma_{i}(x)\Gamma_{j}(y)$, where $\Gamma_{0}(x)$ and $\Gamma_{2}(x)$ are, respectively, the localized ground and second excited states of a one--dimensional unperturbed optical lattice site.
The respective energies of $\ket{ij}$ are $E_{ij}=\hbar \omega_{ij}$, where $E_{00}<E_{02}=E_{20}<E_{22}$.
Clearly the lattice must be deep enough to support this many bound states. The number of bound states in one dimension is plotted against $V_{0}$ in Fig. \ref{fig_bound}. In this paper, we consider a different physical operation, namely amplitude modulation, than the shaking examined in \cite{kiely_2016}.
This leads to a different driving Hamiltonian $H_1 (t)$.

We assume that  $f_x(t)=g_x(t)\cos(\omega_{x}t)$ where the amplitude $g_{x}(t)$ varies slowly relative to $\cos(\omega_{x}t)$. Moreover, the fast oscillations should be done 
on resonance with the transition $\ket{00} \to \ket{20}$ and so $\omega_x=\omega_d \equiv \omega_{20} - \omega_{00}$. After neglecting fast-oscillating terms, we arrive at the following four--level Hamiltonian
\begin{eqnarray}
H_{4L}(t)&=&\frac{\hbar}{2}[\Omega^{(1)}_{x}(t)\ket{20}\bra{00} -\Omega_{c}(t)\ket{02}\bra{20} \nonumber \\
               &+&\Omega^{(2)}_{x}(t)\ket{02}\bra{22}+h.c.],
\label{H_4L}
\end{eqnarray}
where the relevant Rabi frequencies are
\begin{eqnarray}
\Omega^{(1)}_{x}(t)&=&\frac{g_{x}(t)\gamma_{0}}{\hbar} G_{2,0,0,0}(t),\nonumber \\
\Omega^{(2)}_{x}(t)&=&\frac{g_{x}(t)\gamma_{0}}{\hbar} G_{0,2,2,2}(t),\nonumber \\
\Omega_{c}(t)&=& \frac{2 V_{c}(t) \gamma_{1}}{\hbar}.
\end{eqnarray}
The full derivation and technical details, as well as the definitions of
$\gamma_0$,$\gamma_1$ and $G_{n,m,p,q}(t)$ can be found in Appendix \ref{app1}.
It is clear from this result that the state $\ket{22}$ can not be neglected and should be included in the approximation,
as it is resonantly coupled to $\ket{02}$.

The validity of the rotating wave and slowly--varying envelope approximations
can be heuristically combined in the single condition $T \gg \omega_{d}^{-1} \approx (2\omega)^{-1}$ where $\omega=\sqrt{2V_0k^{2}/m}$ is the frequency of the harmonic approximation. The effectiveness of these approximations will be checked in the next section by comparing with the numerical integration of the full Schr\"odinger equation.

\subsection{Initial and target states}

Our goal is to perform a state transfer from the ground state $\ket{00}$ to the angular momentum state
\begin{eqnarray}
\ket{+}=\frac{1}{\sqrt{2}}\left(\ket{20}+ i\ket{02}\right) .
\end{eqnarray}
In the harmonic limit, $L_{z} \ket{+}=2 \hbar \ket{+}$ where  $L_{z}$ is the $z$ component of the angular momentum operator.

We can see that $H_1(t)$ is the same at every lattice site. This is apparent since the term is invariant under the lattice shift operations $x \rightarrow x\pm 2 \ell$ and $y \rightarrow y\pm 2 \ell$. This ensures the ferromagnetic pattern shown in Fig. \ref{fig_setup_diagram}.

Note that if one were interested in creating angular momentum states in an alternating or checkerboard pattern (similar to the one in \cite{kiely_2016}), one could choose a longer wavelength $\lambda_{s}=\sqrt{2} \lambda$, so that $\Omega_{c}$ would alternate sign at every lattice site.

\section{Sequential scheme for preparing an angular momentum state\label{schemes}}

\begin{figure}[t]
\begin{center}
\includegraphics[angle=0,width=0.95\linewidth]{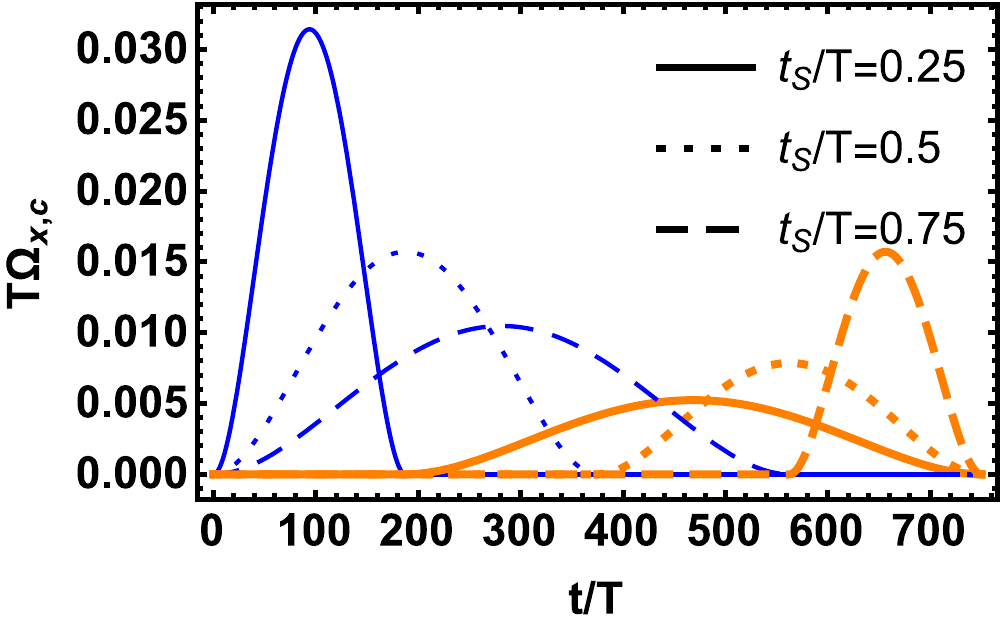}

\end{center}
\caption{Rabi frequencies against time for different values of $t_{S}/T$: $\Omega_{x}$ (blue thin lines) and $\Omega_c$ (orange thick lines). \label{fig_schemes}}  
\end{figure}

In this section, we present a sequential scheme which allows us to prepare our target state in the four-level approximation, i.e. $\Omega_x (t)$ and $\Omega_c (t)$.
By construction, this scheme will give fidelity one exactly in the four-level approximation.
We then convert the effective couplings $\Omega_x (t)$ and $\Omega_c (t)$ back to the
physical quantities: oscillation of the primary lattice amplitude in the $x$ direction, $f_x(t)$, and the amplitude of the rotated lattice, $V_{c}(t)$.
This will allow us to verify if the scheme also works in the full Schr\"odinger equation with high fidelity.

\subsection{Scheme in the four-level approximation}

The idea is first to performs a $\pi$ pulse in $\Omega_{x}$ (of duration $t_S$) which transfers all the population from $\ket{00}$ to $\ket{20}$, followed by a $-\pi/2$ pulse in $\Omega_c$ (of duration $T-t_S$) which leads to the superposition $\ket{+}$.
This method also has the advantage that the state $\ket{22}$ is never populated, which reduces the loss of population to higher levels. 

If we are using sequential pulses (i.e. if either $g_x$ or $V_c$ is non-zero, then the other must be zero) the Rabi frequencies simplify to
\begin{eqnarray}
\Omega_{x}=\Omega^{(1)}_{x}=\Omega^{(2)}_{x}=\frac{g_{x}\gamma_{0}}{\hbar},
\end{eqnarray}
where we assume $g_x(t)$ is first implemented and only afterwards is $V_c(t)$ performed.
The amplitudes of the Rabi frequencies are determined by the switch time $t_S$ and are given by (see Fig. \ref{fig_schemes}),
\begin{eqnarray}
\Omega_{x}(t)&=& \begin{cases} 
\frac{30 \pi  t^2 (t-t_S)^2}{t_S^5}  & 0\leq t \leq t_S , \\ 0 & t_S < t\leq T ,
   \end{cases}\nonumber \\
\Omega_c(t)&=&\begin{cases} 
      0 & 0\leq t < t_S  ,\\
      \frac{15 \pi (t - T)^2 (t - t_S)^2}{(T - t_S)^5} & t_S\leq t \leq T .
\end{cases}
\label{omegapiece}
\end{eqnarray}
Note that  $\Omega_{x}$, $\Omega_c$ and their respective derivative are zero at the start and the end of the process. They also fulfill $\int_{0}^{T}\Omega_{x}(t)dt=\pi$ and $\int_{0}^{T}\Omega_c(t)dt=\pi/2$. Using a square envelope would be problematic due to its broad Fourier spectrum (i.e. the approximation that $g_x$ is slowly varying would not be fulfilled).

\subsection{Numerical simulations of the sequential scheme\label{Numerics}}

\begin{figure}[t]
\begin{center}
\includegraphics[angle=0,width=0.95\linewidth]{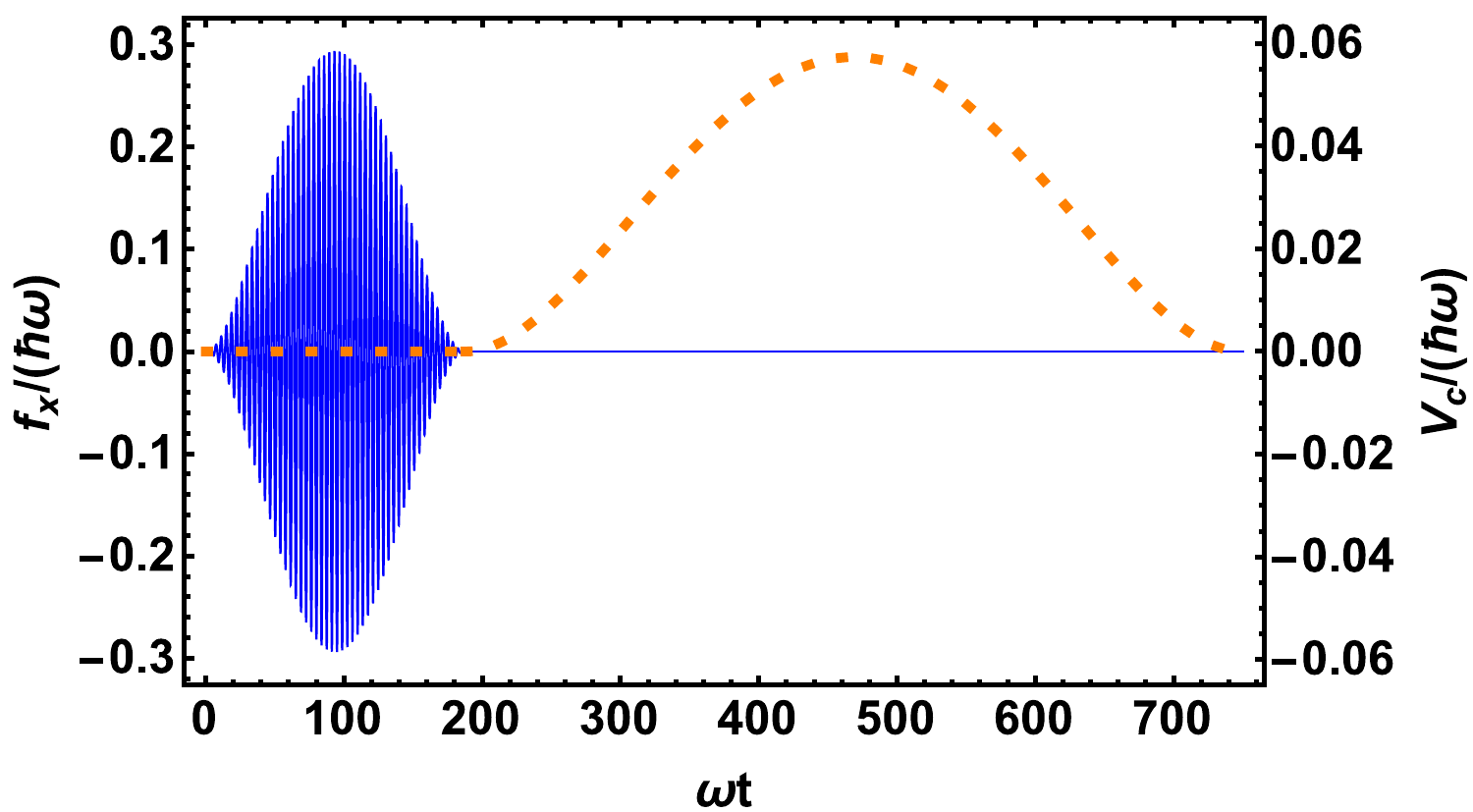}
\end{center}
\caption{Amplitude modulation $f_x(t)$ with $\omega_{x}=\omega_{d}$ (thin, blue solid line) and amplitude of additional lattice $V_{c}(t)$ (thick, orange dotted line) versus time for $t_S=0.25T$, $V_0 = 3\hbar\omega$ and $T=750 \omega^{-1}$. \label{real_fkt}}  
\end{figure}

In order to verify the approximations used to derive this model, we now simulate of the full Schr\"{o}dinger equation with Hamiltonian Eq. \eqref{H_lattice_frame} in coordinate space for an atom initially in the ground state of a single lattice site. The first step is to translate the coupling coefficients $\Omega_x (t)$ and $\Omega_x (t)$ in the four-level approximations back to the physical control parameters, $f_x(t)$ and $V_{c}(t)$.
They relate to the Rabi frequencies as
\begin{eqnarray}
f_x(t) &=& \frac{\hbar}{\gamma_{0}} \Omega_x (t) \cos\left(\omega_{d}t\right) \label{map1} ,  \\
V_{c}(t) &=&\frac{\hbar \Omega_{c}(t)}{2 \gamma_{1}} \label{map2},
\end{eqnarray}
in the sequential case. An example of the resulting functions for the process is shown in Fig. \ref{real_fkt}. The required strength of amplitude modulation is only a fraction of the unperturbed lattice amplitude $V_{0}$.

\begin{figure*}[t]
\begin{center}
(a)\includegraphics[angle=0,width=0.45\linewidth]{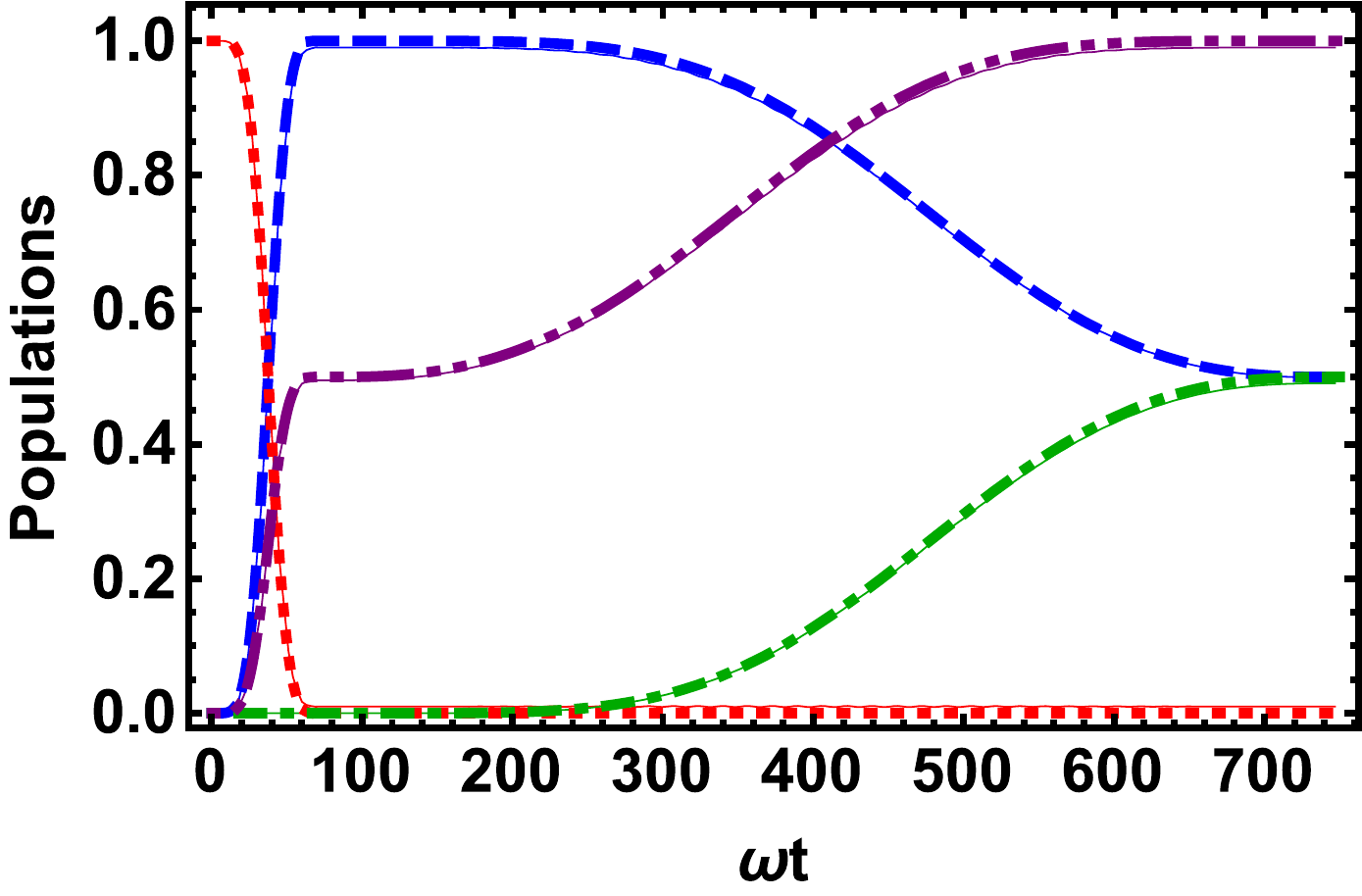}
\hspace{0.5cm}
(b)\includegraphics[angle=0,width=0.45\linewidth]{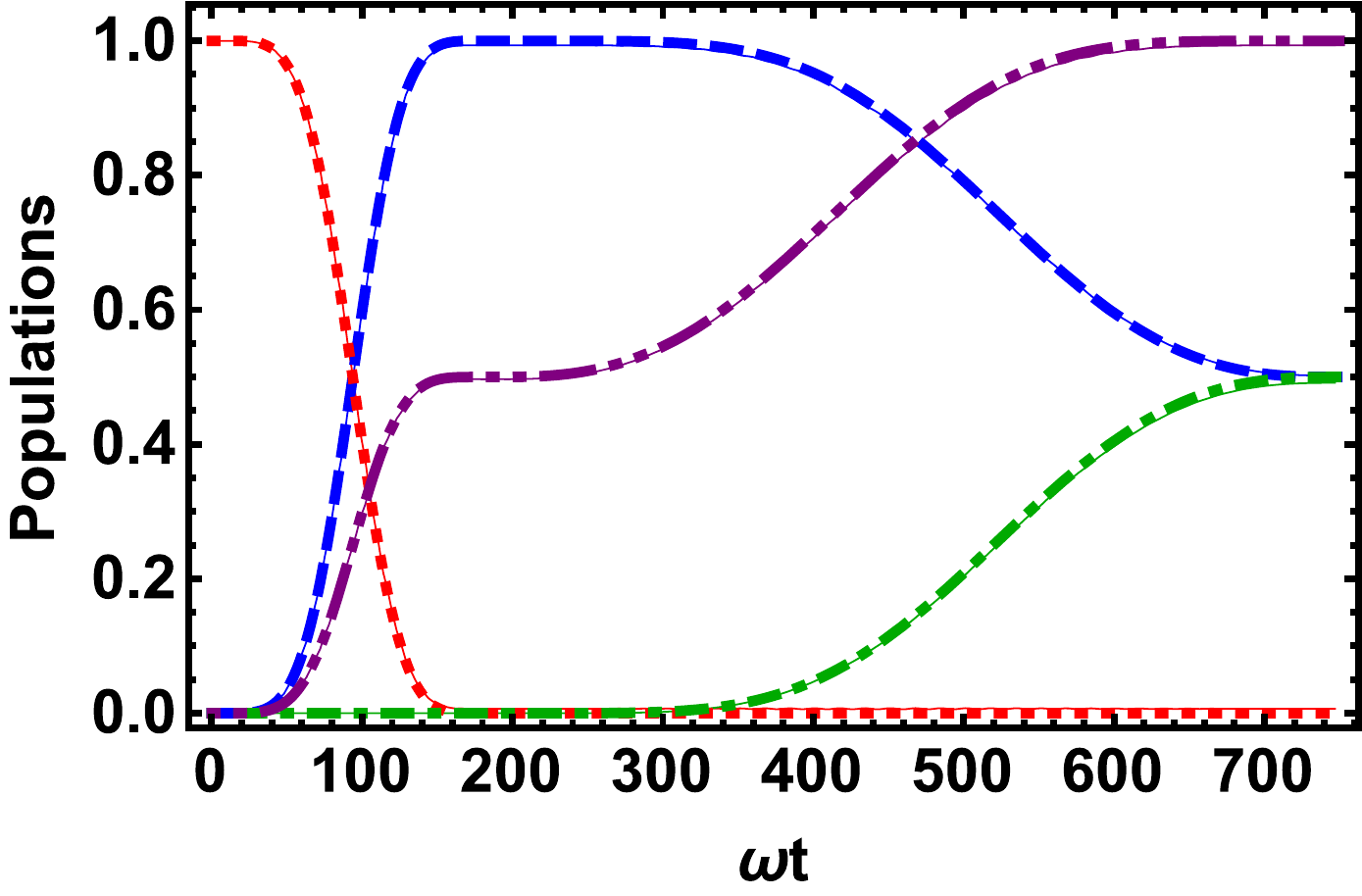}
\\[1cm]

(c)\includegraphics[angle=0,width=0.45\linewidth]{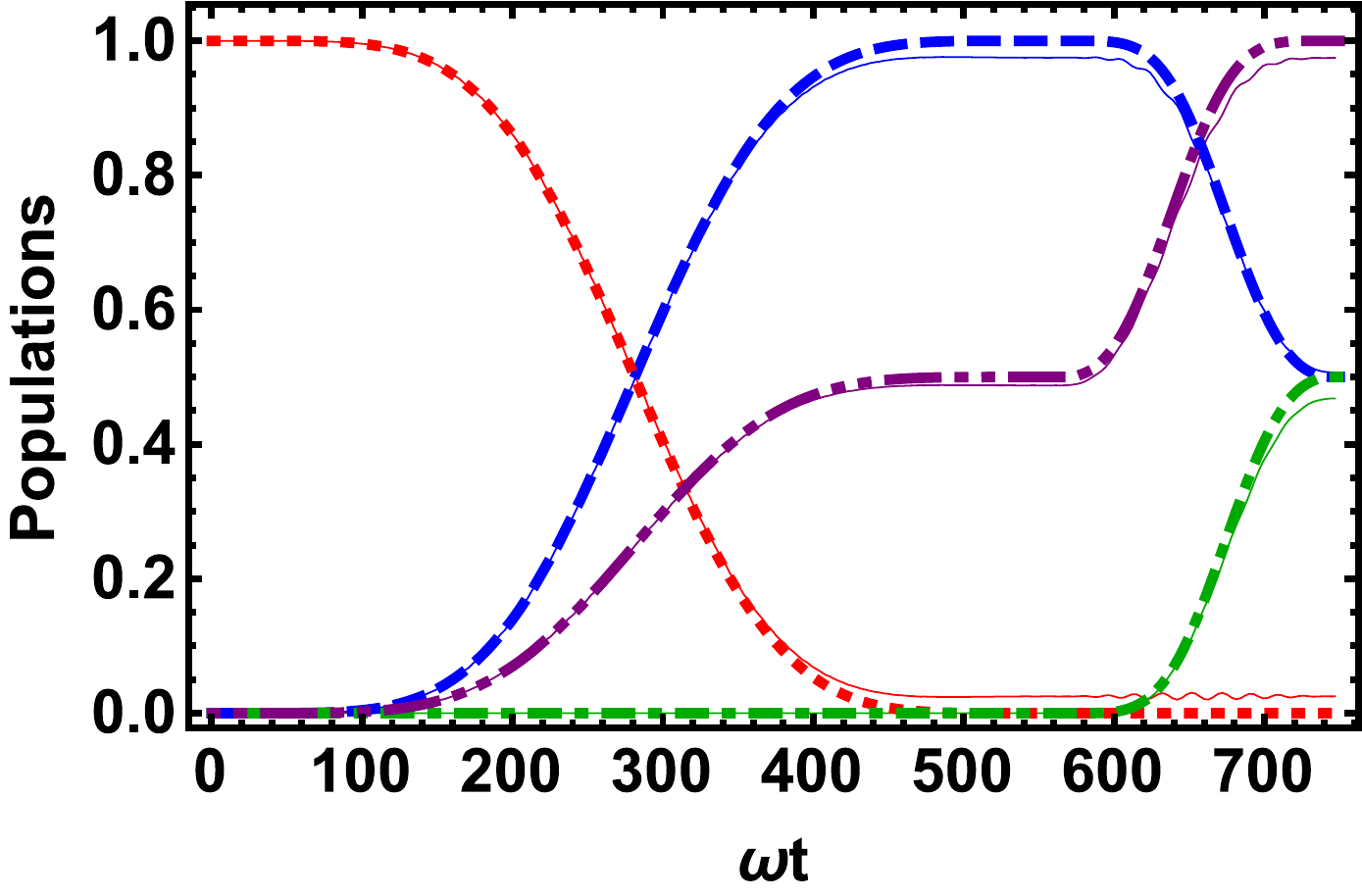}
\hspace{0.5cm}
(d)\includegraphics[angle=0,width=0.45\linewidth]{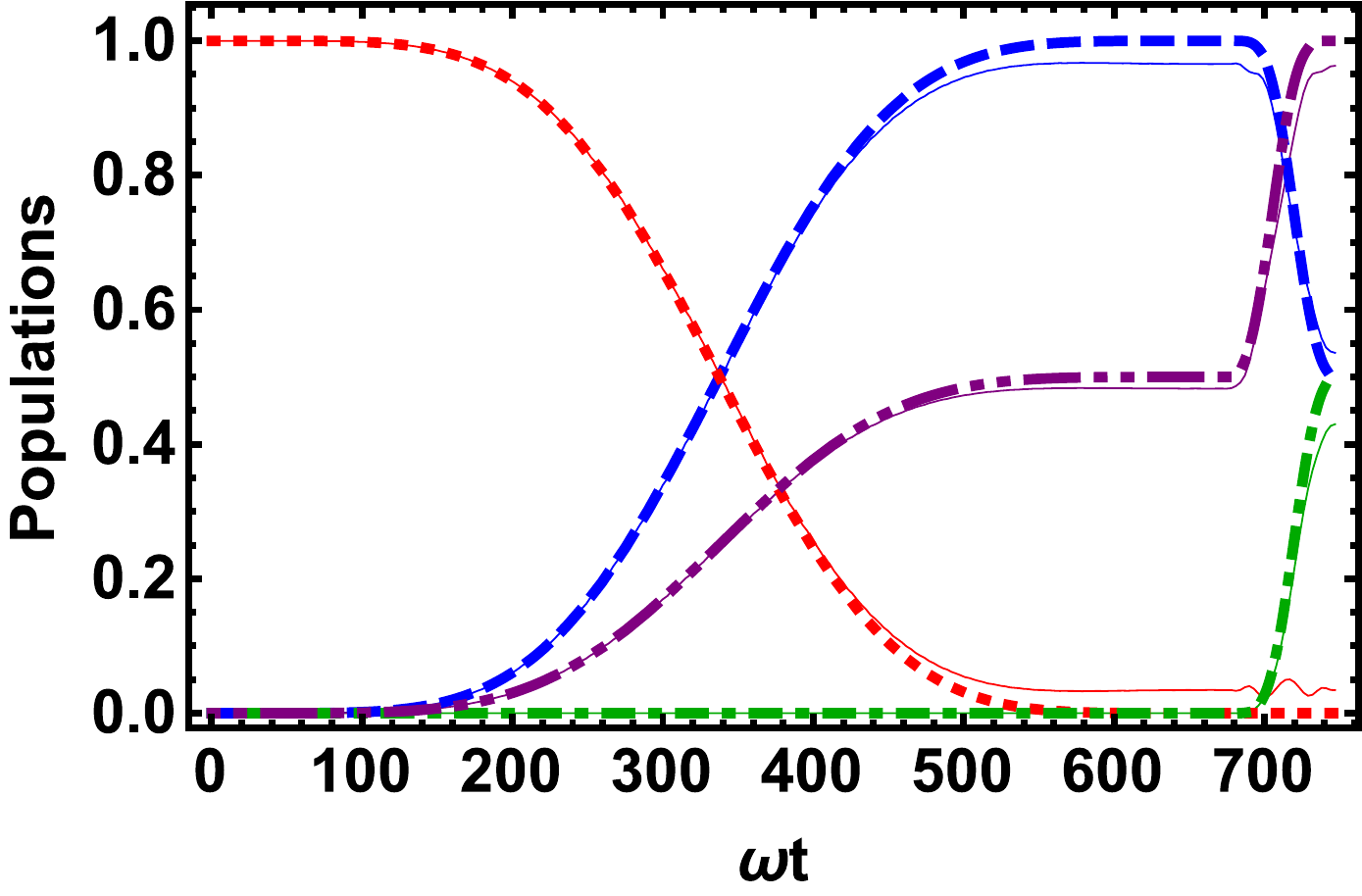}

\end{center}
\caption{Populations against time for $V_0=3 \hbar \omega$ and $T=750 \omega^{-1}$. Dynamics using the four--level approximation (broken lines) and the full Schr\"{o}dinger equation (thin solid nearby lines):
  $\left|\braket{\psi(t)}{00}\right|^{2}$ (red dotted),
  $\left|\braket{\psi(t)}{10}\right|^{2}$ (blue dashed),
  $\left|\braket{\psi(t)}{02}\right|^{2}$ (green dot-dashed), and
  $\left|\braket{\psi(t)}{+}\right|^{2}$ (purple dot-dot-dashed).
  (a) $t_{S}/T=0.1$ (b) $t_{S}/T=0.25$ (c) $t_{S}/T=0.75$ (d) $t_{S}/T=0.9$. \label{populations}}  
\end{figure*}

The time evolution of the Schr\"odinger equation is performed by means of the Fourier split--operator method \cite{split_op}, where the initial ground state is found by imaginary--time evolution. We restrict our simulations to the dynamics of an atom in a single well since we have assumed the Mott--insulator regime.


The results of the numerical simulations of the scheme for several values of $t_{S}$ are shown in Fig. \ref{populations}, together with the ideal populations based on the four--level Hamiltonian in Eq. \eqref{H_4L}. Each subfigure corresponds to a different switch time $t_S$ and a fixed total time $T=750 \omega^{-1}$. The thin, solid lines correspond to the full Schr\"odinger equation and the broken lines correspond to the four level approximation. In all the subfigures, one can see the two distinct steps of the process for both cases. First there is the population inversion between states $\ket{00}$(red dotted line) and $\ket{20} $(blue dashed line). After which there is a $\pi/2$ -pulse between states $\ket{20}$(blue dashed line) and $\ket{02}$ (green dot-dashed line) leading to the superposition state $\ket{+}$ (purple dot-dot dashed line).

Note that during the whole process the maximum population leakage is minimal ($<0.02$ for all subfigures) and the four level approximation accurately reproduces the population dynamics of the full Schr\"odinger equation, not just the final state. The population of state $\ket{22}$ is $0$ throughout the whole process for the four-level approximation as one would expect (Sec. \ref{four_level_approx}). However it is also effectively zero ($<10^{-6}$) for the full Schr\"odinger equation.

The fidelity of the full Schr\"odinger equation leads to final fidelities greater than $0.96$ regardless of the value of $t_S$. This confirms that the mapping to the four--level model is accurate and the scheme works correctly. Some values of $t_S$ do produce higher fidelities than others. Notably $t_S/T=0.9$ (see Fig. \ref{populations}(d)) has the worst final fidelity while $t_S/T=0.25$ (see Fig. \ref{populations}(b)) has the best.

In Fig. \ref{populations}(b), there is good agreement between the approximation and the full dynamics. While this agreement is not as good in Fig. \ref{populations}(d), this is clearly not due to population leakage. The connection between the Rabi frequencies and physical control parameters (see Eqs. \eqref{map1} and \eqref{map2}) becomes less valid here leading to imperfect population inversion.

\begin{figure*}[t]
\begin{center}
(a)\includegraphics[angle=0,width=0.45\linewidth]{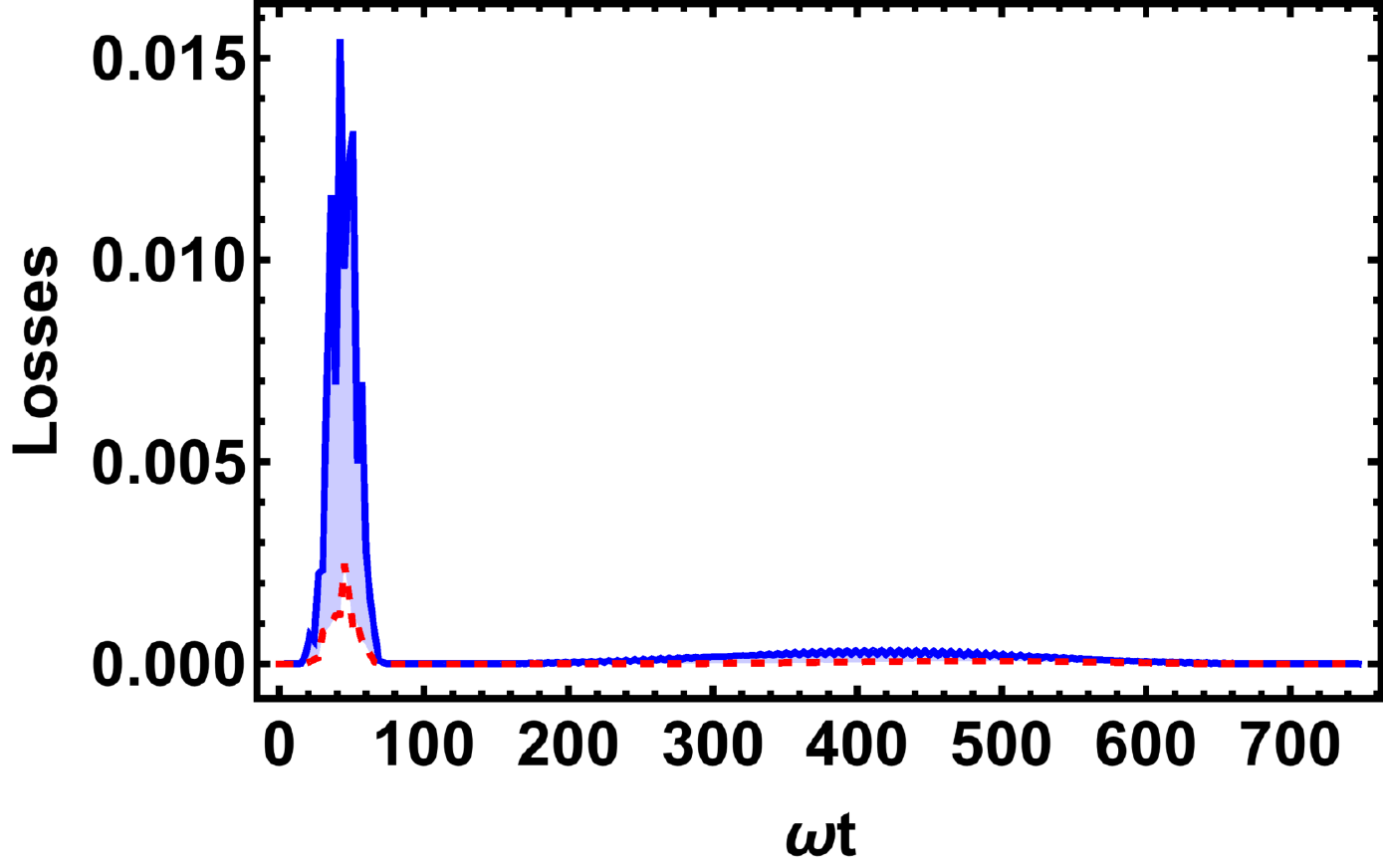}
\hspace{0.5cm}
(b)\includegraphics[angle=0,width=0.45\linewidth]{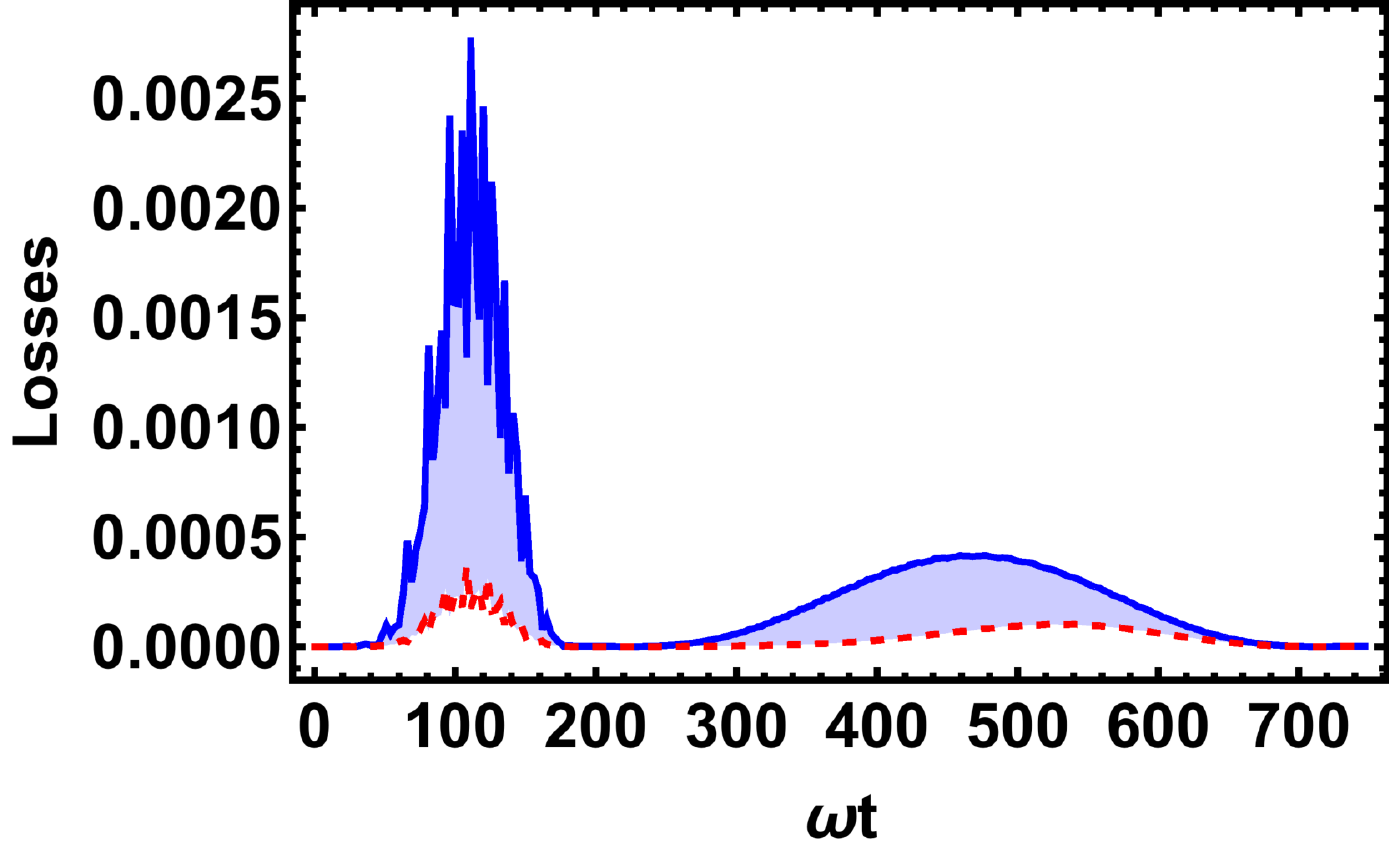}
\\[1cm]

(c)\includegraphics[angle=0,width=0.45\linewidth]{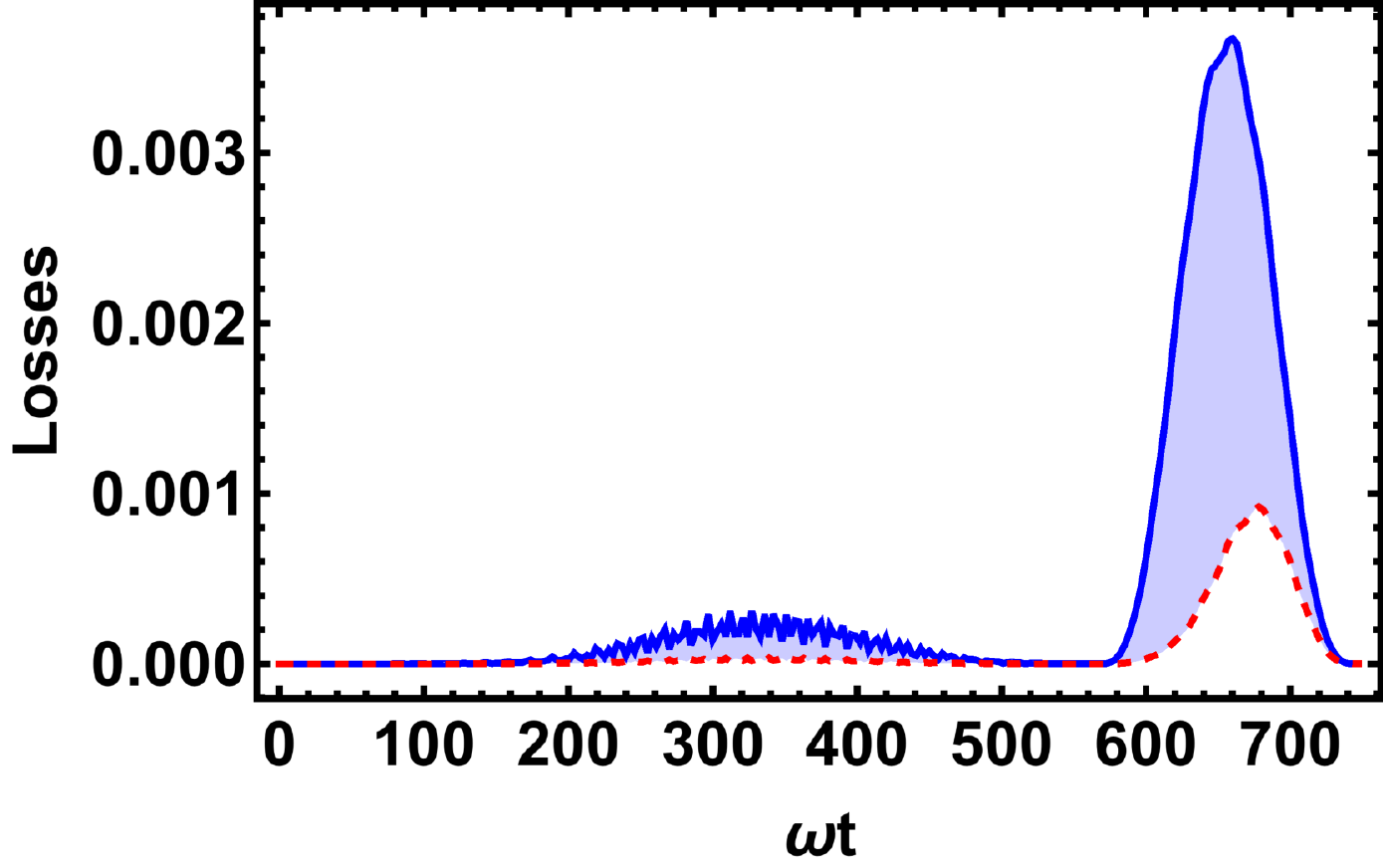}
\hspace{0.5cm}
(d)\includegraphics[angle=0,width=0.45\linewidth]{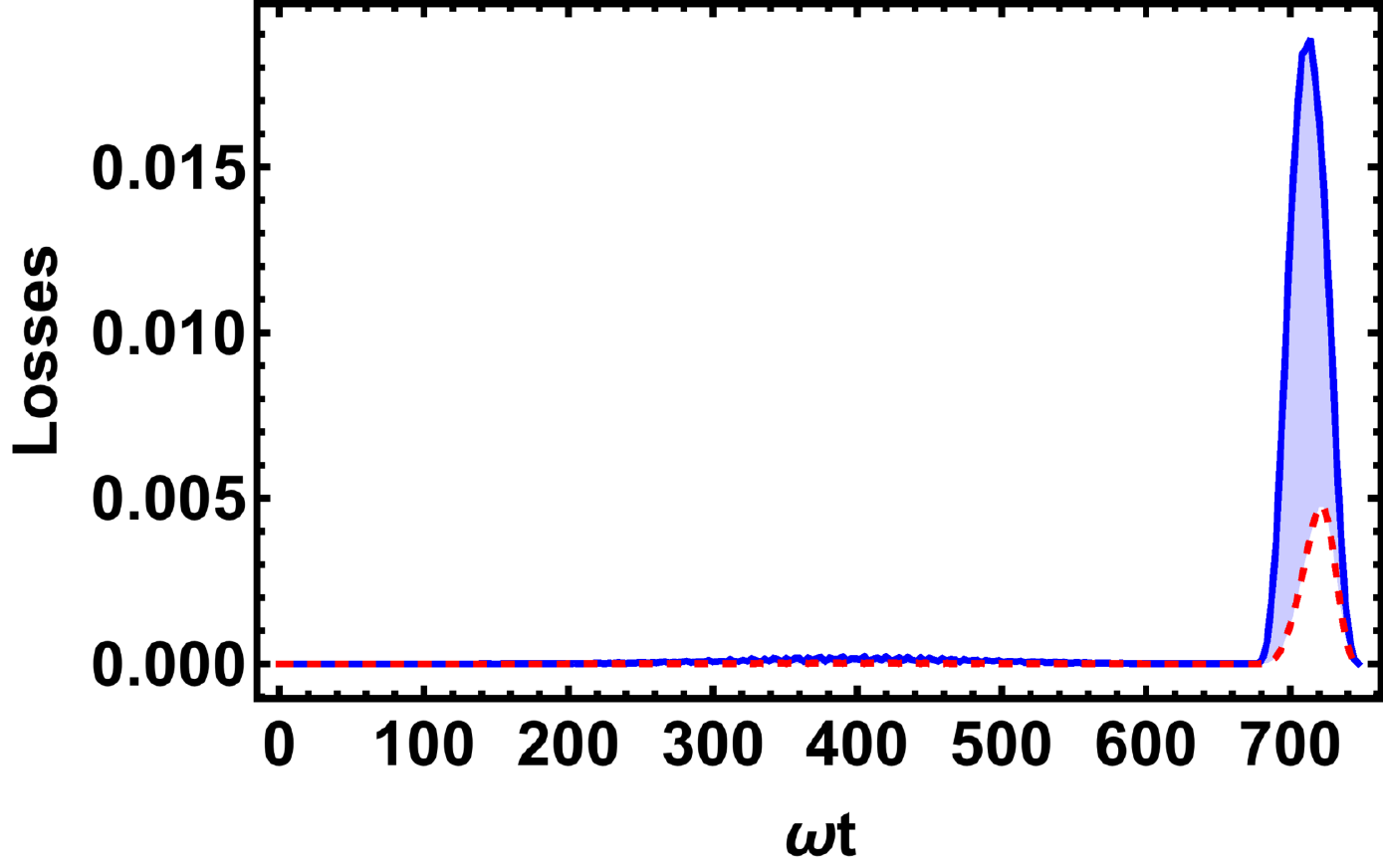}

\end{center}
\caption{Population losses outside the subspace against time for $V_0=3 \hbar \omega$ and $T=750 \omega^{-1}$;  $1-\sum_{i,j \in \{0,2\}}\left|\braket{\psi(t)}{ij}\right|^{2}$ (blue solid upper line), $1-\sum_{i,j \in \{0,2\}}\left|\braket{\psi(t)}{ij}\right|^{2}-\left|\braket{\psi(t)}{40}\right|^{2}$ (red dashed lower line) and $\left|\braket{\psi(t)}{40}\right|^{2}$ (blue shaded area)  (a) $t_{S}/T=0.1$ (b) $t_{S}/T=0.25$ (c) $t_{S}/T=0.75$ (d) $t_{S}/T=0.9$. \label{pop_losses}}  
\end{figure*}

Of course this model is never perfectly valid, leading to population losses which can be seen in Fig. \ref{pop_losses}. Different switch times $t_S$ are shown in the subfigures while the total time is the same in all. The setting shown corresponds to the previous figure (Fig. \ref{populations}).

Even though the total losses outside the four dimensional subspace at the final time are extremely small, it is still useful to identify the most critical source of errors. The states $\ket{40}$ and $\ket{04}$ are the lowest energy states of the correct parity which are neglected in the four-level approximation. Due to the path chosen (i.e. oscillating in $x$ rather than $y$ in the initial step), the most dominant source of losses/leakage is to the state $\ket{40}$.

Therefore apart from the total loss (blue solid lines), the loss into any state other than $\ket{40}$(red dashed line) is also shown in Figs. \ref{pop_losses} such that the shaded blue region correspond to the loss into state $\ket{40}$. In Figs. \ref{pop_losses} (a) and (b), the main loss is during the first step to state $\ket{40}$ (blue shaded area). One can see the oscillations of this loss which originate from the oscillations $f_x(t)$.

In Fig. \ref{populations}(d), there is an imperfect population inversion. However the loss (see Fig.  \ref{pop_losses}(d)) during this phase is negligible.
This underlines that this infidelity is not due to leakage to other levels but to the imperfect population inversion originating from the mapping between $\Omega_x$ and the coupling strength $f_x(t)$. Even in the second step it can be seen in Figs. \ref{pop_losses}(b)-(d), that the state $\ket{40}$ is still the most relevant.

Note that the maximum unwanted excitations occur at the maximum intensities of the two sequential pulses and there are higher losses for pulses of shorter duration.

\begin{figure}[t]
\begin{center}
\includegraphics[angle=0,width=0.9\linewidth]{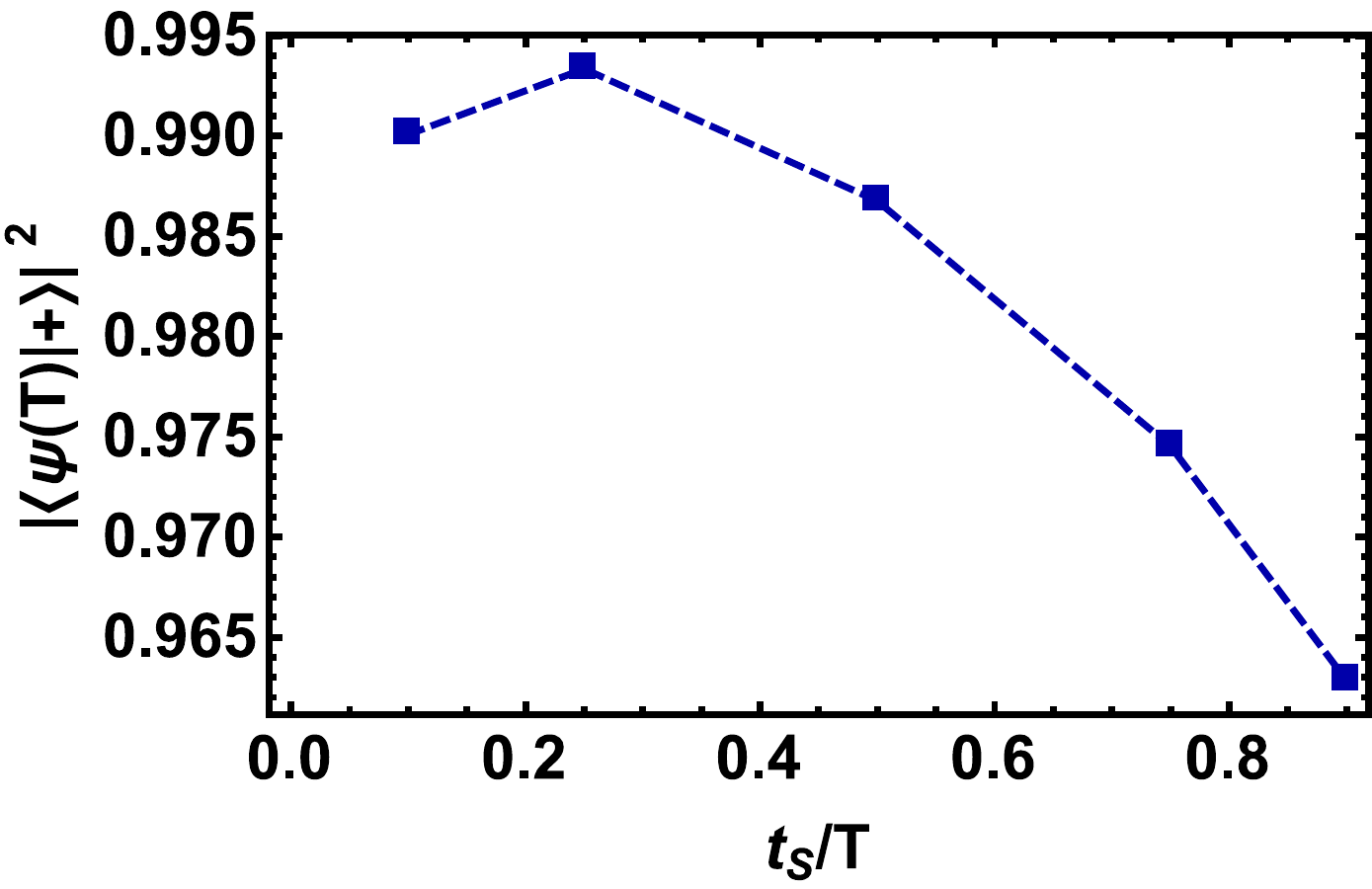}
\end{center}
\caption{Fidelity $\left|\braket{\psi(T)}{+}\right|^{2}$ against $t_{S}$ for $V_0 = 3\hbar\omega$ and $T=750 \omega^{-1}$. \label{ts_plot}}  
\end{figure}

\subsection{Fidelity dependence on different physical parameters}

We now consider how the fidelity of this scheme depends on different physical parameters.
First, the final fidelity of the scheme for different switch times $t_S$ is shown in Fig. \ref{ts_plot}.
The highest fidelity is obtained for a switch time of $t_{S}/T \approx 0.25$. This effect is likely due to the fact that the second pulse must neglect many more transitions in the rotating wave approximation than the first pulse (see Eq. \eqref{preRW}) and hence would require more operation time. As the choice of $t_{S}$ does not affect the fidelity greatly, from this point on we will fix $t_{S}/T=0.25$.

\begin{figure}[t]
\begin{center}
\includegraphics[angle=0,width=0.9\linewidth]{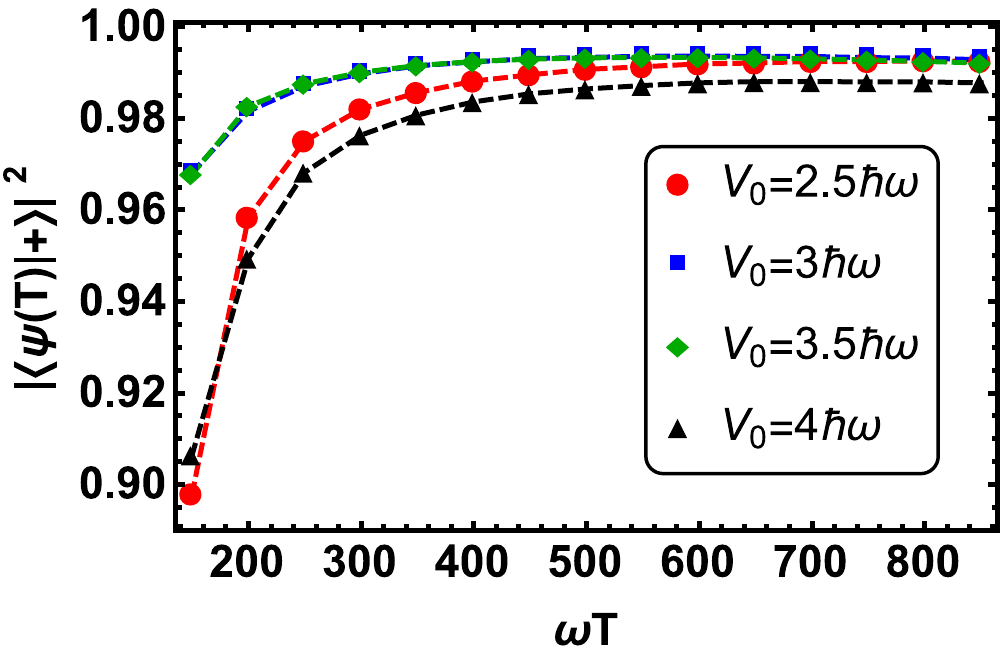}
\end{center}
\caption{Fidelity $\left|\braket{\psi(T)}{+}\right|^2$ against total time $T$ for different lattice depths $V_{0}$ for a fixed trapping frequency $\omega$ with $t_S=0.25T$.
Points joined with lines:
$V_0=2.5\hbar\omega$ (red circles),
$V_0=3.0\hbar\omega$ (blue squares),
$V_0=3 .5\hbar\omega$ (green diamonds) and
$V_0=4.0\hbar\omega$ (black triangles).
\label{fidelity_plots}
}  
\end{figure}

In Fig. \ref{fidelity_plots}, we can see the fidelity for different total times $T$ and different lattice depths $V_{0}$. As expected, the fidelity generally increases as the total time $T$ increases, since the rotating wave approximation becomes more valid in this regime. This highlights that the four--level model breaks down for very short operation times. The lattice depth also slightly effects the fidelity, with the maximum fidelities achieved for $V_{0} \approx 3\hbar \omega $. For very shallow depths the target state is likely too weakly bound and close to the continuum. However for large lattice depth, the energy levels become equally spaced and other states cannot be neglected. This heuristically explains why the optimal depth is this intermediate value, since the four--level model does not account for the effect of all these other levels.

\begin{figure}[t]
\begin{center}
\includegraphics[angle=0,width=0.9\linewidth]{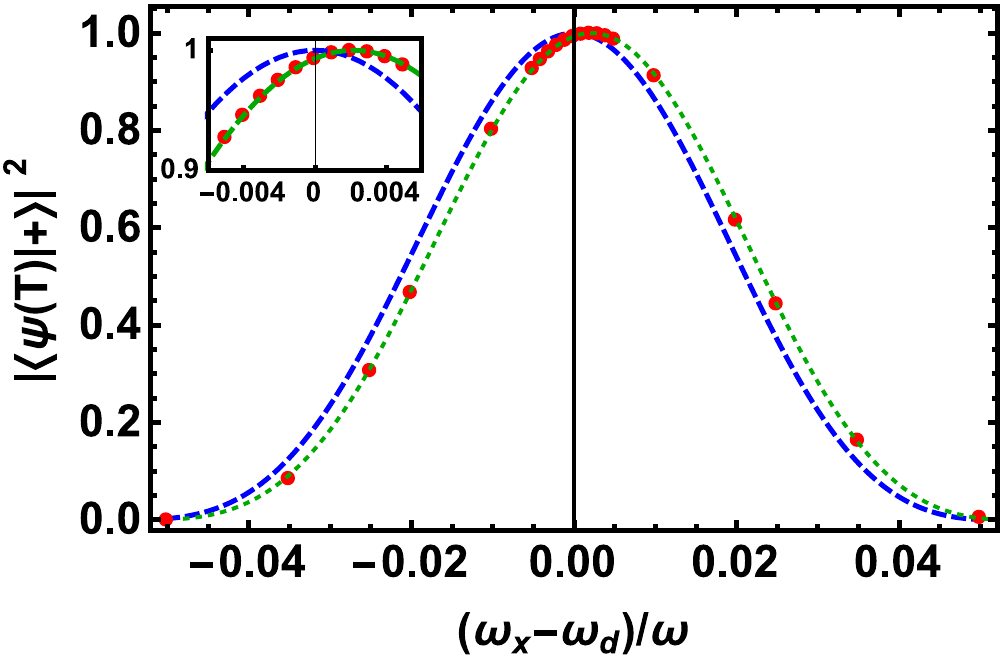}
\end{center}
\caption{Fidelity $\left|\braket{\psi(T)}{+}\right|^2$ against the deviation from resonant oscillation $(\omega_{x}-\omega_{d})/\omega$ for $V_0=3\hbar\omega$, $T=750\omega^{-1}$ and $t_S=0.25T$ .
Red points correspond to the full Schr\"{o}dinger equation, dashed blue line to the four--level model and dotted green line to the four--level model shifted by $\approx 0.0021$.
\label{resonance_peak}}  
\end{figure}

Finally, in Fig. \ref{resonance_peak}, we can see the resonance curve for the processes, i.e., the fidelity against the detuning of the amplitude modulation frequency. We compare the four--level model after applying the rotating wave approximation but without assuming $\omega_{x}=\omega_{d}$, against the full Schr\"{o}dinger equation dynamics. As expected, one achieves high fidelity when the amplitude oscillation frequency is essentially on resonance. One can see that the process is highly selective (full width at half max $\approx 0.0427 \omega^{-1}$). 

By assuming a constant Rabi frequency and considering the detuned transitioned transition
between $\ket{00}$ and $\ket{20}$, one can obtain an explicit formula for the fidelity as a function of
detuning. It roughly varies as shifted $\sinc^2(x)=\sin^2(x)/x^2$.
Motivated by this, we have fit our data with this curve and obtain an R squared value of $0.99997$. Other typical resonance functions such as Gaussian, Lorentzian or Voigt do not provide as good a fit. 
Hence this resonance curve is most accurately modeled by a $\sinc^2(x)$ function. Since the second pulse is not affected by using a different frequency
amplitude modulation, this effect is not the result of multiple transitions.

However the highest fidelity of the full dynamics is achieved for a slightly off resonant frequency $\omega_{x} \approx \omega_{d}+0.0021 \omega$. This is not true in the four--level model, as the corresponding curves have their maximum at resonance. The reason for this shift is the presence of an off resonant coupling to the state $\ket{40}$ (which is not present in the four-level model). By slightly increasing the detuning of $\Omega_{x}$ with respect to the $\ket{00} \leftrightarrow \ket{20}$ transition, an even greater detuning in the coupling between $\ket{20}$ and $\ket{40}$ is created, leading to less leakage to these higher states.

In detail, this can be seen explicitly by adiabatically eliminating $\ket{40}$, which adds a detuning term. Note that $E_{40}<E_{22}=2E_{20}$ which implies that $\hbar \omega_d> E_{40}-(E_{20}+E_{00})$ leading to a positive detuning for the state $\ket{40}$.
Therefore the adiabatic elimination leads to an effective, positive detuning acting on state $\ket{20}$, the positive shifts the value of $\omega_x-\omega_d$ results in a negative detuning on state $\ket{20}$ and the maximum fidelity corresponds roughly to a cancelation of these two detunings. Shifting the four-level model results by $0.0021$ (green dotted line) corresponds very well with the results from the full dynamics. Similar effects can be seen in \cite{kiely_2016,cao_2017}.


\section{Experimental considerations \label{exp_sec}}

The optical potential in Eq. \eqref{Vpot} could be implemented in a number of ways. 
%
%
It can be implemented by superimposing two square optical lattice potentials with wavelengths that differ by a factor of $2$. Since one is rotated with respect to the other, the corresponding required lattice geometry is achieved; this
has been experimentally shown in \cite{Liberto_2006} and references therein.
%
%

An alternative way is by shining two laser beams of the same wavelength at an angle to generate the required a one-dimensional lattice where
the well distance can be adjusted by changing the angle (see \cite{Kruger_2007} for an experimental implementation of this). This basic idea
to generate one-dimensional lattices can be generalised using an additional pair of lasers at a right angle to the first pair to generate
a two-dimensional optical lattice with the required effective wavelength.

%
%
Another such possibility is optical tweezers, where there are a variety of established techniques.
Acousto-optic deflectors allow one to control the position and intensity of a laser beam. An acoustic wave generates a defractive pattern for the laser leading to arbitrary two-dimensional atomic arrays \cite{barredo_2016}.
Another example are liquid crystal spatial light modulators \cite{nogrette_2014} which can imprint a specific phase pattern on the laser beam being focused by a lens. In such a way the intensity profile in the focal place is the Fourier transformation of this phase-modified beam.
%
Digital mirror devices, which are arrays of micro-mechanical mirrors, allow holographically generation of arrays of dipole traps \cite{stuart_2018}.
This device imprints a binary (mirrors can be ``on" or ``off") amplitude hologram of the desired trapping potential on the beam. This is then transformed on to an asperic lens and the trapping potential is again formed in the focal plane of the lens.
With such a variety of techniques available, optical tweezers could prove to be a useful alternative implementation.
%
%


A state of atoms with non-zero orbital angular momentum can be detected by measuring the density-density correlation function \cite{liu_2008}. Parameter values of $V_0/(\hbar \omega)=3.5$ and $\omega T=500$ could for example be experimentally realized using $^{133}$Cs atoms with lasers of wavelength $\lambda=1064 \, \text{nm}$ and a lattice depth of $49 E_{r}$ for the unperturbed lattice, where $E_{r}=\frac{\hbar^{2}k^{2}}{2 m}$ is the recoil energy.
The amplitude oscillation frequency required would be $\omega_{d}/(2\pi) \approx 31\,\text{kHz}$ and the total operation time would be $T \approx 4.3 \, \text{ms}$.

We have estimated the tunneling frequency by simulating the central atom alone on a $3 \times 3$ lattice. The tunneling rate for the second excited state between two sites is given by $R_2 \approx 0.00157 \omega \approx 183 \,\text{Hz}$ for $V_0/(\hbar \omega)=3.5$. The time scale associated with this is $1/R_{2} \approx 5.5 \, \text{ms}$ which is longer than the operation time $T$. This could also have been approximated by
\begin{equation}
R_2\approx \frac{2}{\hbar} \int_{-\ell}^{3 \ell} \Gamma_{2}(x) V_{0} \sin^{2}(k x) \Gamma_{2}(x-2 \ell) dx.
\end{equation}
Note that the effective tunneling rate for the atom during the process is overestimated here since the natural repulsive interaction between the atoms is not accounted for and $R_2$ is assumed to be the relevant tunneling rate during the whole process. While the ground state tunneling rate $R_0$ is also relevant, $R_2>R_0$ so $R_2$ provides the strictest estimate of the operation time needed.

\section{Conclusions \label{discuss}}

We have developed a scheme to prepare a large angular momentum state, namely one with each atom having approximately two units of orbital angular momentum,
starting from a Mott insulator state in an optical lattice. This is done by modulating the lattice amplitude and the addition of a rotated rectangular lattice.


The methods proposed here could be used in conjunction with the results from \cite{kiely_2016} to form a modular system (or building blocks) for creating different higher orbital states. Each particular operation fulfills a different general role. 

This work can be extended by using non-sequential pulses designed using Lewis-Riesenfeld invariants \cite{LR69} for four--level systems \cite{kiely_2016,Li2018}. Designing the pulses in this way would have the important advantage that they could be optimized against noise, systematic errors or unwanted transitions to higher levels \cite{ruschhaupt_2012, kiely_2014}. The four--level model derived is quite general and could be used to prepare other superpositions of the basis states.

Since these results are for the Mott insulator regime (where only one atom populates each potential well), they could also be useful in single atom optical tweezers experiments where one can achieve $ \sim 90\%$ single atom occupancy in such a trap or collection of traps \cite{Carpentier_2013}. This would be an attractive option for studying two atom interactions in the context of angular momentum.

\section*{Acknowledgements}
We would like to thank David Gu\'ery-Odelin for providing insightful comments on the manuscript. This work was supported by the Basque Country Government (Grant No. IT986-16), and MINECO/FEDER,UE (Grant No. FIS2015-67161- P).

\begin{appendix}
\section{Derivation of the Four--level Approximation \label{app1}}

In this appendix we will provide the full derivation of the four--level model in Eq. \eqref{H_4L}.
Let $\tilde H (t) = \sum_{j=00,02,20,22}\sum_{k=00,02,20,22} \ket{j} \bra{j}H(t)\ket{k}\bra{k}$.
We want to remove most of the the diagonal terms of $\tilde H(t)$. Therefore, we define a unitary transformation of the form
\begin{eqnarray}
U(t)&=& e^{i( \omega_{x}-\omega_{20})t}\chi_{00}(t)\ket{00}\bra{00}+e^{-i \omega_{20}t}\chi_{02}(t)\ket{02}\bra{02}
\nonumber \\ &+&e^{-i \omega_{20}t}\chi_{20}(t)\ket{20}\bra{20}+e^{-i \omega_{22}t}\chi_{22}(t)\ket{22}\bra{22} ,\nonumber \\
\end{eqnarray}
under which the Hamiltonian changes as $H \rightarrow U^{\dagger}HU-i \hbar U^{\dagger}\dot{U}  = H_{4L}$. Note that the unperturbed lattice is separable which gives $\omega_{22}=2\omega_{20}-\omega_{00}$. This leads to 
\begin{eqnarray}
H_{4L}&=&\hbar(\omega_{x}-\omega_{d})\ket{00}\bra{00} \nonumber \\
&+&\left[\gamma_{0} f_{x}(t) -V_{c}(t) \gamma_{2} \right]e^{i \omega_{x} t}\tilde{\chi}_{2,0,0,0}(t) \ket{20}\bra{00}\nonumber \\
&+&\left[\gamma_{0} f_{x}(t)-V_{c}(t) \gamma_{3} \right] e^{-i\omega_{d}t} \tilde{\chi}_{0,2,2,2}(t) \ket{02}\bra{22}\nonumber \\
&-&V_{c}(t) \gamma_{1} e^{-i(\omega_{x}+\omega_{d})t}\tilde{\chi}_{0,0,2,2}(t) \ket{00}\bra{22}\nonumber \\
&-&V_{c}(t) \gamma_{2} e^{-i\omega_{x}t} \tilde{\chi}_{0,0,0,2}(t) \ket{00}\bra{02}
\nonumber \\
&-&V_{c}(t) \gamma_{1} \tilde{\chi}_{2,0,0,2}(t) \ket{20}\bra{02}\nonumber \\
&-&V_{c}(t) \gamma_{3} e^{-i\omega_{d}t}\tilde{\chi}_{2,0,2,2}(t) \ket{20}\bra{22}\nonumber \\
&+&\textrm{h.c.},
\label{preRW}
\end{eqnarray}
where we have defined
\begin{eqnarray}
\alpha_{n} &=& \int_{-\ell}^{\ell} \Gamma_{n}^{2}(x)\sin^{2}(k x) dx,\\
\beta_{n} &=& \int_{-\ell}^{\ell} \Gamma_{n}^{2}(x)\cos(2 k x) dx, \\
\chi_{n,m}(t)&=& \exp \left \{ -\frac{i}{\hbar} \left[\alpha_{n} \int_{0}^{t} ds f_{x}(s)-\beta_{n}\beta_{m}\int_{0}^{t}ds V_{c}(s)\right]\right \}.\nonumber \\
\end{eqnarray}
and
\begin{eqnarray}
\tilde{\chi}_{n,m,p,q}(t)&=&\chi_{n,m}^{*}(t)\chi_{p,q}(t), \\
\gamma_{0} &=&\int_{-\ell}^{\ell} \Gamma_{0}(x)\sin^{2}(k x)\Gamma_{2}(x)dx , \\
\gamma_{1} &=&\left[\int_{-\ell}^{\ell} \Gamma_{0}(x)\cos(2 k x)\Gamma_{2}(x)dx\right]^{2} , \\
\gamma_{2} &=&\int_{-\ell}^{\ell} \Gamma_{0}(x)\cos(2 k x)\Gamma_{0}(x)dx \nonumber \\
&\times&  \int_{-\ell}^{\ell} \Gamma_{0}(y)\cos(2 k y)\Gamma_{2}(y)dy, \\
\gamma_{3} &=&\int_{-\ell}^{\ell} \Gamma_{0}(x)\cos(2 k x)\Gamma_{2}(x)dx \nonumber \\
&\times& \int_{-\ell}^{\ell} \Gamma_{2}(y)\cos(2 k y)\Gamma_{2}(y)dy.
\end{eqnarray}

\begin{figure}[t]
(a)\includegraphics[angle=0,width=0.95\linewidth]{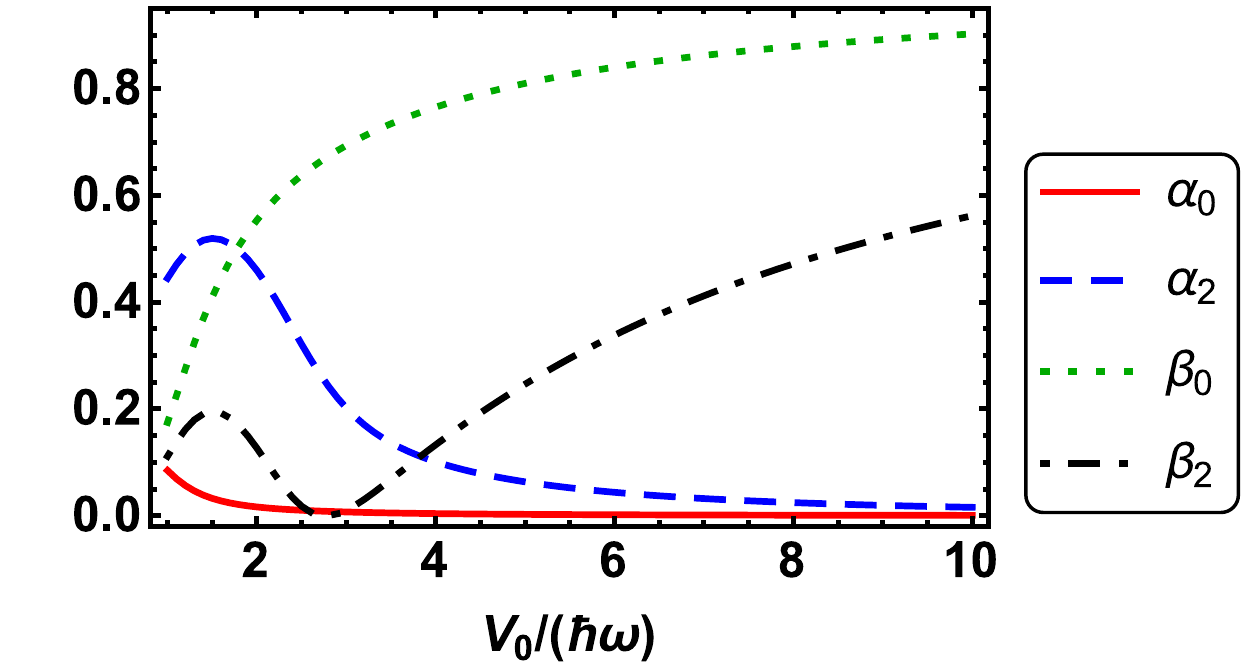}\\ 
(b)\includegraphics[angle=0,width=0.95\linewidth]{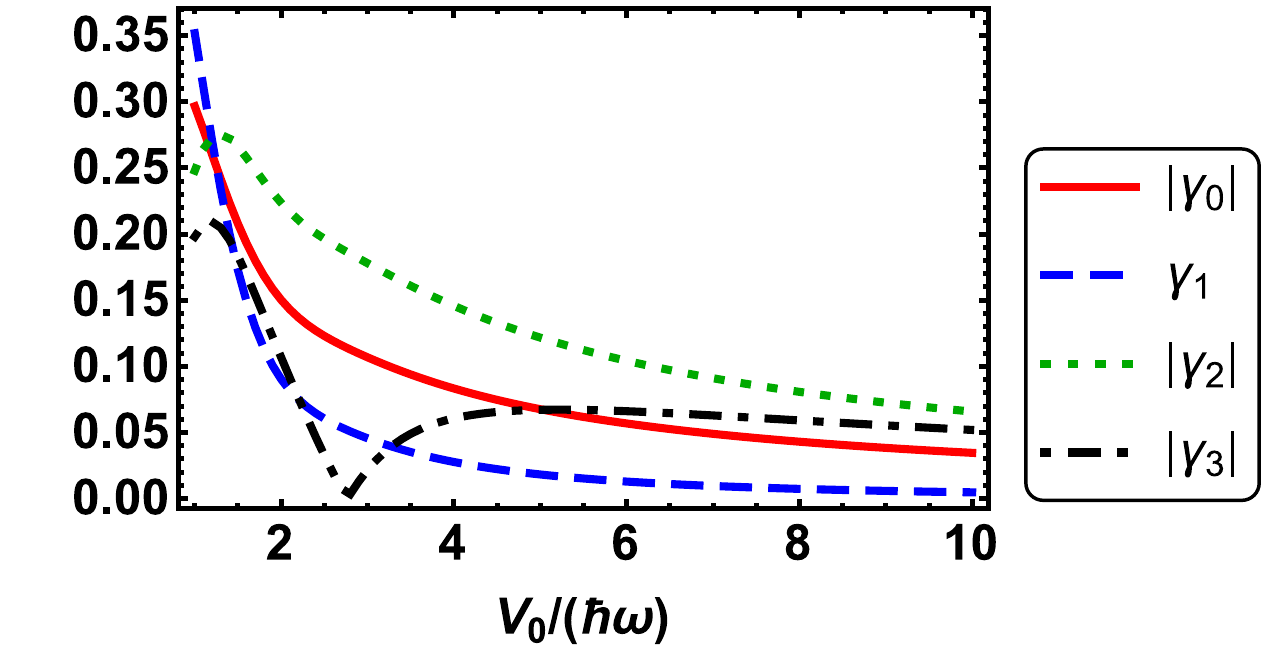}\\ 
(c)\includegraphics[angle=0,width=0.85\linewidth]{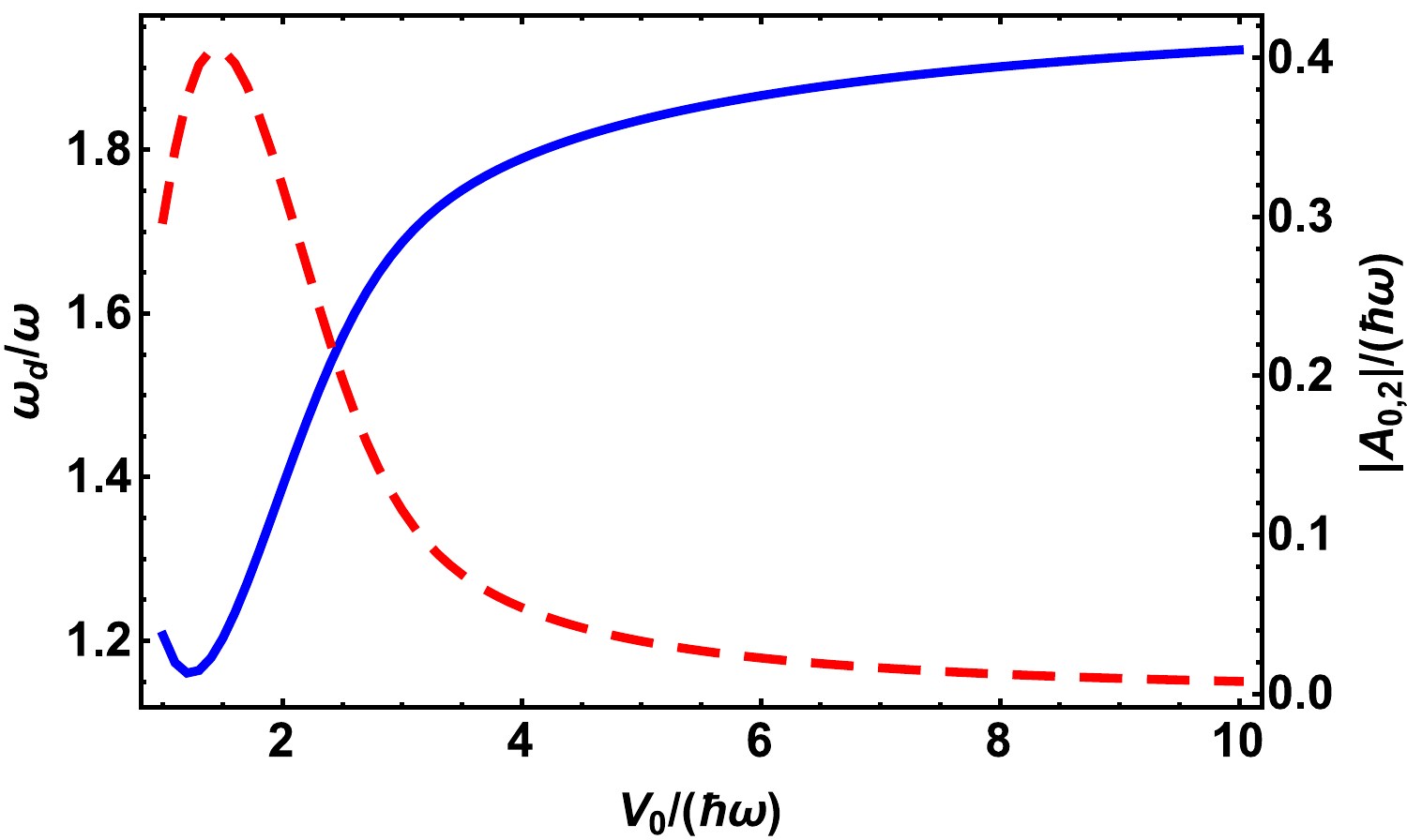}

\caption{Parameters (a) $\alpha_{n}$ and $\beta_{n}$, (b) $\gamma_{n}$, and 
(c) $\omega_{d}$ (solid blue line) and $A_{0,2}$ (dashed red line) against lattice depth $V_{0}$.
 \label{fig_param}}  
\end{figure}

The parameters $\alpha_{0,2}$ and $\beta_{0,2}$ (which determine to what extent some terms can be neglected) are plotted for different values of $V_{0}$ in Fig. \ref{fig_param}(a). In the harmonic limit ($V_0\rightarrow\infty$), $\alpha_{0,2}\rightarrow0$ and $\beta_{0,2}\rightarrow 1$. The parameters $\gamma_n$ (which determine how strongly states are coupled) are shown in Fig. \ref{fig_param}(b) where one can clearly see that $\gamma_{n}\rightarrow 0 \, \forall n$ in the harmonic limit.

We assume that $f_{x}(t)=g_{x}(t)\cos(\omega_{x}t)$, i.e. it consists of a slowly varying envelope $g_{x}(t)$ and a fast oscillating term $\cos\left(\omega_{x}t\right)$ with $\omega_{x}=\omega_{d}$. This resonant frequency, $\omega_d \rightarrow 2\omega$ in the harmonic limit and is shown in Fig. \ref{fig_param}(c). 

This allows us to simplify the $\tilde{\chi}_{n,m,p,q}(t)$ terms. Firstly using partial integration we have
\begin{eqnarray}
\int_{0}^{t} f_{x}(s) ds &=& \int_{0}^{t} g_{x}(s) \cos(\omega_{d}s) ds \nonumber \\
									   &=& \frac{1}{\omega_{d}}\left[g_{x}(t) \sin(\omega_{d}t)-\int_{0}^{t} \dot{g}_{x}(s) \sin(\omega_{d}s) ds\right]	\nonumber \\
									   &\approx & 	\frac{1}{\omega_{d}}g_{x}(t) \sin(\omega_{d}t).		  
\end{eqnarray}
Secondly we make use of the Jacobi-Anger expansion \cite{abramowitz1983},
\begin{eqnarray}
e^{-i \kappa \sin (\Omega t)}=\sum_{k=-\infty}^{\infty} J_{k}(\kappa)e^{-i k \Omega t}
\end{eqnarray}
where $J_{k}(\kappa)$ is a Bessel function of the first kind and $\kappa$ is constant in time. We assume that this relation is also approximately valid for $\kappa$ slowly varying relative to a fast oscillating $\sin (\Omega t)$ term.
If we now define $A_{p,n}=\frac{\alpha_{p}-\alpha_{n}}{\hbar \omega_{d}}$, we can write the first type of term in $H_{4L}$ (see Eq. \eqref{preRW}) as
\begin{eqnarray}
&& e^{\pm i \omega_{d}t}\gamma_{0} f_{x}(t)\tilde{\chi}_{n,m,p,q} \nonumber \\
&=& \frac{g_{x}(t)}{2} \gamma_{0}\left(1+e^{\pm 2 i \omega_{d}t}\right) \tilde{\chi}_{n,m,p,q}\nonumber \\
&\approx& \frac{g_{x}(t)}{2}\gamma_{0} \left\{\sum_{k=-\infty}^{\infty} J_{k}\left[A_{p,n}g_{x}(t)\right]\left[e^{-i k \omega_{d}t}+e^{-i(k\mp 2) \omega_{d}t}\right]\right\}  \nonumber \\
&\times& G_{n,m,p,q}(t) \nonumber \\
&\approx& \frac{g_{x}(t)}{2}\gamma_{0}\left\{J_{0}\left[A_{p,n}g_{x}(t)\right]+J_{\pm 2}\left[A_{p,n}g_{x}(t)\right]\right\} G_{n,m,p,q}(t), \nonumber \\
\end{eqnarray}
where in the last step we have assumed that all fast rotating terms can be ignored (i.e. a rotating wave approximation) and used the definition
\begin{eqnarray}
G_{n,m,p,q}(t)=\exp\left[\frac{i}{\hbar}(\beta_{p}\beta_{q}-\beta_{n}\beta_{m})\int_{0}^{t}ds V_{c}(s)\right].
\end{eqnarray}

Note that $G_{n,m,p,q}(t)$ survives the rotating wave approximation since $\beta_{p}\beta_{q}-\beta_{n}\beta_{m} \ll \omega_{d}/\omega$.

The second type of term in $H_{4L}$ is given for $a\in \{1,2,3\}$ and $b\in \{0,1,2\}$ as
\begin{eqnarray}
&& V_{c} \gamma_{a} e^{\pm b i \omega_{d}t}\tilde{\chi}_{n,m,p,q} \nonumber \\
&\approx& V_{c} \gamma_{a} e^{\pm b i \omega_{d}t} \left \{ \sum_{k=-\infty}^{\infty} J_{k}\left[A_{p,n}g_{x}(t)\right] e^{-i k \omega_{d}t} \right \} G_{n,m,p,q}(t) \nonumber \\
&\approx& V_{c} \gamma_{a} J_{\pm b}\left[A_{p,n}g_{x}(t)\right], \nonumber 
\end{eqnarray}
where in the last step we have again made a rotating wave approximation.

In order to get the desired coupling structure, we use the fact that $\fabs{A_{p,n}}\ll 1$. This is easy to see since $A_{0,0}=A_{2,2}=0$ and $\fabs{A_{2,0}}=\fabs{A_{0,2}} \ll 1$ (see Fig. \ref{fig_param}(c)). Using this approximation we set $J_{0}(A_{0,2}g_{x}) \approx 1$ and  $J_{1,2}(A_{0,2}g_{x}) \approx 0$.
After making these last approximations, one arrives at the four--level model Hamiltonian in Eq. \eqref{H_4L}.

To summarize, we have used the following approximations in this derivation: there are only four relevant basis states, the function $g_{x}(t)$ varies slowly relative to $\cos(\omega_{d}t)$, i.e.,  $\fabs{\int_{0}^{t} \dot{g}_{x}(s) \sin(\omega_{d}s) ds} \ll 1$, and $A_{0,2}\ll 1$.

\end{appendix}

\end{document}